\documentclass[10pt,twocolumn,a4paper]{article}

\usepackage[a4paper, left=2cm, right=2cm, bottom=3.5cm]{geometry}
\usepackage[pdftex]{graphicx}
\graphicspath{{figures/}}
\usepackage[utf8]{inputenc}
\usepackage{url}
\usepackage{xspace}
\usepackage{hyperref}
\usepackage{color}
\usepackage{mathptmx}

\usepackage{subcaption}	

\newcommand{\name}{{\scshape Palpatine}\xspace}
\newcommand{\titlename}{Palpatine\xspace}
\newcommand{\footurl}[1]{\footnote{\url{#1}}}

\newcommand{\spara}[1]{\smallskip\noindent\textbf{#1}}

\newenvironment {squishlist}
{\begin{list}{$\bullet$}
  { \setlength{\itemsep}{0pt}
     \setlength{\parsep}{3pt}
     \setlength{\topsep}{3pt}
     \setlength{\partopsep}{0pt}
     \setlength{\leftmargin}{1.5em}
     \setlength{\labelwidth}{1em}
     \setlength{\labelsep}{0.5em} } }
{\end{list}}

\date{}

\begin{document}

\title{\titlename: Mining Frequent Sequences for Data Prefetching in NoSQL Distributed Key-Value Stores}

\author{Sérgio Esteves, João Nuno Silva, and Luís Veiga \\
INESC-ID, Instituto Superior Técnico, Universidade de Lisboa
}



%


\maketitle

\begin{abstract}
This paper presents \name, the first in-memory application-level cache for Distributed Key-Value (DKV) data stores, capable of prefetching data that is likely to be accessed in an immediate future. To predict data accesses, \name continuously captures frequent access patterns to the back store by means of data mining techniques. With these patterns, \name builds a stochastic graph of accessed items, and makes prefetching decisions based on it.   

Experimental evaluation indicates that \name can improve the latency of a specific DKV store by more that an order of magnitude.
\end{abstract}


%

\section{Introduction}
\label{sect:introduction}

The colossal volume of data, that is generated on a daily-basis by web-based organizations, is calling for sophisticated models and systems that can take advantage of patterns and implicit relationships, that are often exhibited by data, in order to leverage their performance. One important class of such systems is caching systems, where data traces, containing meaningful relationships, can be used to improve read latency by predicting, and staging from back store, data pieces that are soon going to be requested with great likelihood.

Recently, there has been an accentuated movement from traditional relational databases into the so-called NoSQL. Examples of NoSQL Distributed Key-Value (DKV) data stores include BigTable~\cite{Chang:2006:BDS:1267308.1267323}, Dynamo~\cite{DeCandia:2007:DAH:1294261.1294281}, Apache Cassandra~\cite{Lakshman:2009:CSS:1582716.1582722} and HBase~\cite{hbase2011george}. The main advantages of these DKV stores over typical RDBMS encompass: shared-nothing horizontal scaling, automatic partition of data and replication over many servers, flexibility in changing the structure of records, efficient use of distributed indexes and augmented performance and availability in terms of latency and throughput~\cite{Cattell:2011:SSN:1978915.1978919, Stonebraker:2010:SDV:1721654.1721659,5993686}. Such higher performance and scalability are achieved at the expense of relaxing the ACID properties, or replacing them by BASE properties~\cite{Pritchett:2008:BAA:1394127.1394128}, thus providing a weaker concurrency model. Further, to migrate data from a RDBMS to a DKV store it is usually necessary to resort to data denormalization~\cite{Wei:2008:SDD:1367497.1367535}, which possibly breaks functional dependencies, relations and indexing information between data entities. 

These first level relations, representing join and indexing relationship, are typically handled in DKV stores explicitly at the application level, thereby batching simple read/write operations or through an SQL skin library (such as Phoenix~\cite{phoenix} for HBase). At the data store tier, it is possible to observe such batches of operations and infer relationships between data entities. In addition to batched operations, we may also capture higher level relations through the data access patterns, such as secondary indexing relationship and causal relations.

There is a plethora of real-world examples of non-deterministic high-level relations. For example, in the context of a social network: $(i)$ when visiting a profile page for the first time, more often than not there is a subsequent click on the profile picture to enlarge it; and $(ii)$ new uploaded photos are likely to be followed by user views soon enough. Or in the context of shopping cart web based applications, for example, $(i)$ there are common frequent purchase sequences in a grocery store (e.g., buying cheese and ham together); and $(ii)$ in the summer a frequent relation of items for a fashion store might be selecting in sequence the options \emph{women}, \emph{dress}, and then \emph{casual}, \emph{beach} or \emph{formal}.
Like these examples, there are in general many frequent patterns when accessing data from a back store. This is also specially relevant in Online Transaction Processing (OLTP) workloads, where many relations between different data entities are exposed in the form of transactions.

We claim that by observing and capturing data access patterns to a DKV store it is possible to improve the hit rate of the caches that come coupled with those stores (e.g., the block cache in HBase or the row cache in Cassandra), and thus reduce latency for interactive applications and augmenting throughput for batch computations. In this work, we propose caching items that are predicted to be accessed in a near future according to a history of past observations.

Caching and data prefetching are effective techniques to reduce and hide the latency of accessing data on file systems~\cite{Cao:1995:SIP:223587.223608,Vanderwiel:2000:DPM:358923.358939}. Data prefetching uses data referencing patterns to anticipate cache misses, thereby fetching data in advance from disk to the cache. To be successful, it is necessary that prefetches are ($i$) timely, assuring data is available on cache before it is actually requested; ($ii$) useful, leading to prefetch hits before being replaced and avoiding cache misses; and ($iii$) efficient, by not introducing any significant overhead. Also, prefetching can suffer from side effects such as cache pollution, resulting from prefetching data that was inaccurately predicted to be accessed in a near future, and increased I/O bandwidth requirements, which results from creating more I/O concurrent requests through prefetching.

To trace data access patterns in DKV stores, we rely on well-known data mining algorithms. Data mining techniques have been used extensively to discover and identify patterns in data accesses to web applications~\cite{Facca2005225}, as well as databases~\cite{553155}. With real reference traces, we are thus able to improve cache hit rate by prefetching data that is predicted to be accessed in a close future with a certain confidence degree.

In this paper, we introduce~\name, an in-memory Key-Value cache at the application level, for DKV stores, that is capable of prefetching data that is likely to be accessed in a near future. \name builds, and updates at runtime, a stochastic graph of frequent sequences of accessed data items. Then, based on cache parameters, like size and current churn rate, we select subgraphs of items to be prefetched in sequential order.

\name can be easily coupled with DKV stores, and is made fully transparent to applications. As a concrete instance, we integrated \name with the wide-columnar store HBase. Experimental evaluation indicates that \name can improve the latency of HBase in more than an order of magnitude.

The key contributions from this paper are:

\begin{squishlist}
	\item We present a solution to leverage the performance of back store caching based on the continuous observation of frequent data access patterns. Our solution is unique in the way real time probabilities of accessed items are used to improve hit rate: we make decisions on the fly to trade-off potentially higher probability of a future cache hit with more churn in the cache. Also, our solution is the first one specifically tailored to DKV stores. 
	\item We provide an implementation of our model with \name, that we integrated into the popular DKV store HBase. 
	\item We conduct an extensive evaluation using realistic benchmarks and demonstrate the benefits of our solution.
\end{squishlist}

The remainder of this paper is organized as follows. In the next section we review relevant related work. Section~\ref{sect:data-mining} presents data mining algorithms and their adequacy to our problem. Following, Section~\ref{sect:architecture} describes the architecture and design choices of \name. Then, experimental evaluation takes place in Section \ref{sect:evaluation} and Section~\ref{sect:conclusion} concludes the paper.

\section{Related Work}
\label{sect:related-work}


There has been a vast body of research in cache prefetching, such as in the context of processor memory caching~\cite{Mittal:2016:SRP:2966278.2907071} and web caching (at the browser, proxy, or server level)~\cite{ali2011survey}. Database caching work, which is the focus of this paper, has been less representative however. Database caching can differ from web caching in a fundamental way: database caching has in general considerably less entropy than web caching since the data and spectrum of use cases are more contained per database.

Most work carried out in database prefetching is based on the observation that sequentiality of access abounds in database workloads~\cite{Smith:1978:SPD:320263.320276,Rodriguez-Rosell:1976:EDR:1300774.1302260,Hsu:2001:IRB:383734.383737}. However, such work is typically constrained to the optimization of range scan operations (e.g.,~\cite{Luan2009,903270}), and do not regard probabilistic user/application-specific frequent piped access operations.



More creative and studied work is found in the area of web cache prefetching. In this area, prefetching can be content-based or history-based. Content-based prefetching depends on the analysis of web page contents to find links that are likely to be accessed in subsequent requests~\cite{1277820}. Naturally, such approach does not fit well database caching. 

History-based prefetching can be typically classified in 4 approaches: $(i)$ dependency graph; $(ii)$ Markov model; $(iii)$ cost function; and $(iv)$ data mining. The dependency graph consists of nodes that represent web pages and directed weighted arcs connecting nodes, specifying that a node can be accessed after another with a certain probability. This approach yields a low prediction accuracy, since only pairs of dependencies between 2 web pages are analyzed.


The cost function prefetches web pages based on factors such as the popularity, update rate and lifetime of a page. The success of this approach entails the need of a strong bias in access patterns.

The data mining approach can be typically classified into association rules-based or clustering-based. Association rules-based discovers groups of pages that are commonly accessed together in the same session. Association rules can take into account factors like recency, adjacency, and order of accessed web pages~\cite{Yang:2004:BAB:982721.982726}. A significant drawback is that a great amount of useless rules are generated, due to many patterns observed from all users' references, which results in inaccurate and inefficient predictions.

Cluster-based prefetching discovers groups of similar data instances, called clusters. It can form groups of similar sessions together based on a distance measure between pairs of accessed web pages. In many cases, the number of clusters to be formed needs to be provided in advance. If clusters end up containing a large amount of objects, the corresponding prefetching policy can cause substantial cache pollution.

Algorithms based on data mining and Markov models are the most common in web prefetching, albeit very limited in the literature of database prefetching. Moreover, despite other prefetching techniques being common in relational databases~\cite{Smith:1978:SPD:320263.320276}, they are practically absent in NoSQL DKV stores~\cite{Daniel:2016:PFP:2976767.2976775}, an increasingly popular option nowadays. In this paper, we perform sequential pattern mining on DKV logs in order to build Markov models, thereby exploring unique inherent characteristics of this type of data store, such as dealing with denormalized data.

\section{Data Mining}
\label{sect:data-mining}
As aforementioned, explicit relations in typical relational databases, represented via database schema, turn into implicit relations when moving to NoSQL data stores. Despite such direct relations, we may also observe higher level associations and functional dependencies. We start by describing the data structure of NoSQL data stores, taking as example the HBase, and identifying the different kinds of patterns we may observe there. After that, we discuss existing algorithms and techniques for pattern mining that represent a good fit to our specific problem.

\subsection{Observed patterns in DKV stores}

There are different categories of DKV stores according to their intrinsic data models, such as document, graph, or key-value stores. In this work we target wide-columnar DKV stores, since they are general purpose stores that have been used extensively in recent years. Although abstractions could be provided, we take the HBase data store (the BigTable open-source clone) as an example to better concretize our problem and solution. 

Briefly, HBase is a sparse, multi-dimensional sorted map, indexed by row, column (includes family and qualifier), and timestamp; the mapped values are simply an uninterpreted array of bytes. It is column-oriented, meaning that most queries only involve a few columns in a wide range, thus significantly reducing I/O. Moreover, these databases scale to billions of rows and millions of columns, while ensuring that write and read performance remain constant.

In DKV stores, data is usually accessed and manipulated through guileless call interfaces. These interfaces are also used by SQL skins that convert SQL statements into batches of simple read/write operations. In HBase, \emph{get} and \emph{scan} operations are used to read data.

We identify three main types of access patterns: 
\begin{enumerate}

\item \textbf{Frequent column sequence} corresponds to a sequence of columns (including family and qualifier) that are accessed frequently for the same row. Such sequence might resemble a join in relational databases. For prefetching, this is the most efficient type of pattern, since all data pertaining to a single row is located in the same machine.

\item \textbf{Frequent row sequence} corresponds to a sequence of rows that are accessed frequently for the same columns. It also includes frequent range scans comprising contiguous rows that are accessed frequently for the same columns.

\item \textbf{Frequent hybrid sequence} corresponds to a sequence of items, row and column, that are accessed frequently.
\end{enumerate}


%
%

\spara{Data pre-processing.} Typically, DKV stores do not emit logs of read/write operations due to their potential large dimension. We intercept requests of these operations in \name and we create our own structured backlogs that exempt the system from performing any type of data pre-processing on otherwise external logs, like data integration, cleaning or reduction. 

Our logs consist of user sessions that contain all accessed data containers (extracted from the read requests). A session represents a burst of user activity; i.e., consecutive requests to the datastore where each consecutive pair are not separated by more than a defined time gap. A data container is the metadata that identifies a cell of data in the backstore, and can be a table, row, column (including family and qualifier), or any combination of these.


\subsection{Algorithms for Sequential Pattern Mining}
\label{sect:mining-algos}

\spara{Overview.} There has been a plenitude of algorithms and techniques for finding frequent apriori unknown patterns or associations among elements or events in a given data store~\cite{zhao2003sequential,Mooney:2013:SPM:2431211.2431218,fournier2017survey}. Most existing sequential pattern mining algorithms follow either Apriori-based or Pattern-Growth-based approaches. Apriori-based algorithms perform candidate generation, and can employ either breath-first search (BFS) or depth-first search (DFS) approaches. Pattern-Growth algorithms only consider existing patterns (by repeatedly scanning the database) and employ DFS predominantly.

Sequential pattern mining algorithms can be differentiated through the following factors: efficiency (in terms of time and space complexity), scalability, completeness (finding all proper frequent subsequences), and user-specific constraints. Examples of user-specific constraints include the minimum number of sequences in a database which contain a given subsequence (i.e., minimum support), the minimum and maximum time between two consecutive itemsets in a sequence (i.e., gap), and the maximum time duration for each sequential pattern (i.e., duration).

Sequential pattern mining algorithms have, however, a major limitation: a very large number of patterns may be found (according to the minimum support defined), most of which might not be of great use to the task at hand. Moreover, such large number of patterns makes the performance of algorithms degrade significantly in terms of memory and runtime. To overcome this issue, new algorithms were designed to discover \emph{concise} representations of sequential patterns, which consist of a subset of all sequential patterns that is meaningful and that summarizes the entire set. These algorithms can be orders of magnitude faster than previous traditional algorithms and output a considerable smaller number of patterns. 

There are three main types of concise representations: closed, generator and maximal sequential patterns. Closed are the patterns that are not included in other patterns with the same support; i.e., they represent the largest subsequences extracted from the set of all sequences. Discovering only closed sequential patterns reduces the outputted set of sequences substantially. Generator are the patterns that have no subsequence patterns with the same support. They can be in smaller, equal or larger number than those of closed patterns. Finally, maximal sequential patterns are those that are not strictly included in other closed patterns. The number of maximal patterns can be of several orders of magnitude less than those of closed patterns. However, maximal patterns are not lossless (like closed patterns); i.e., from maximal patterns we can recover all possible sequential patterns and their support without consulting the database.

\begin{figure*}
	\centering
	\begin{subfigure}[t]{0.32\textwidth}
		\centering
	  \includegraphics[width=0.8\columnwidth]{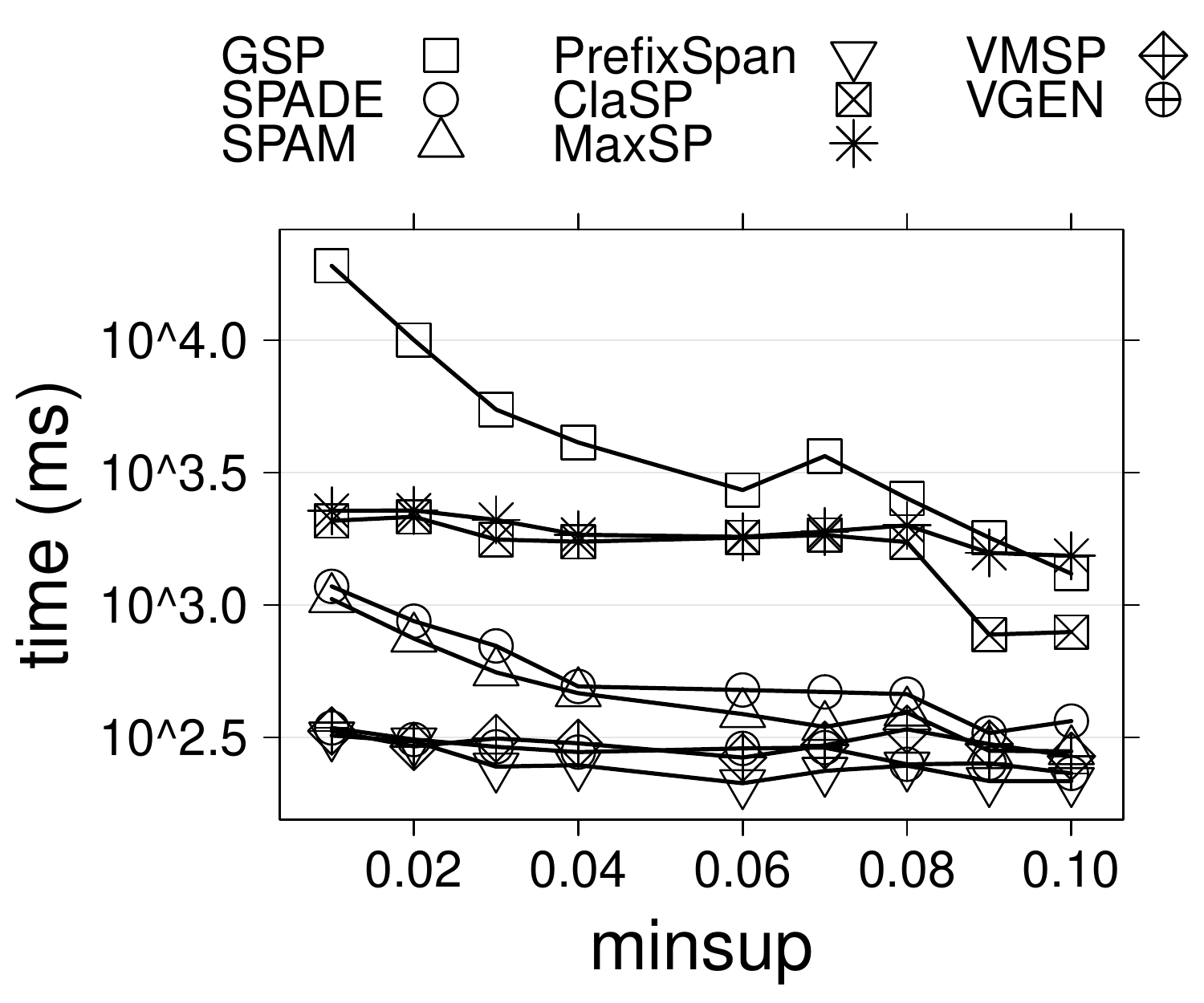}
	\end{subfigure}
	\begin{subfigure}[t]{0.32\textwidth}
			\centering
	  \includegraphics[width=0.8\columnwidth]{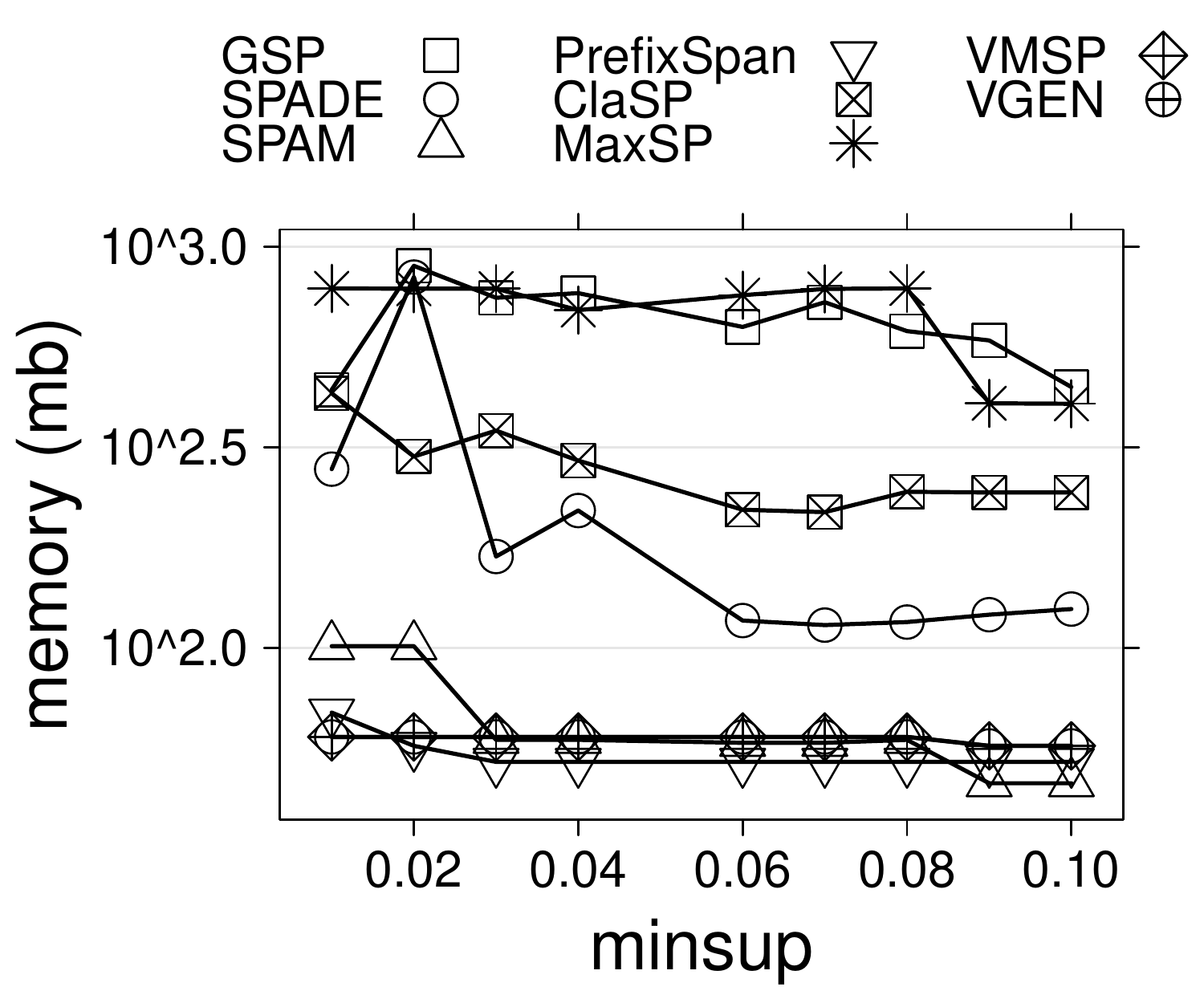}
	\end{subfigure}
	\begin{subfigure}[t]{0.32\textwidth}
		\centering	
		\includegraphics[width=0.8\columnwidth]{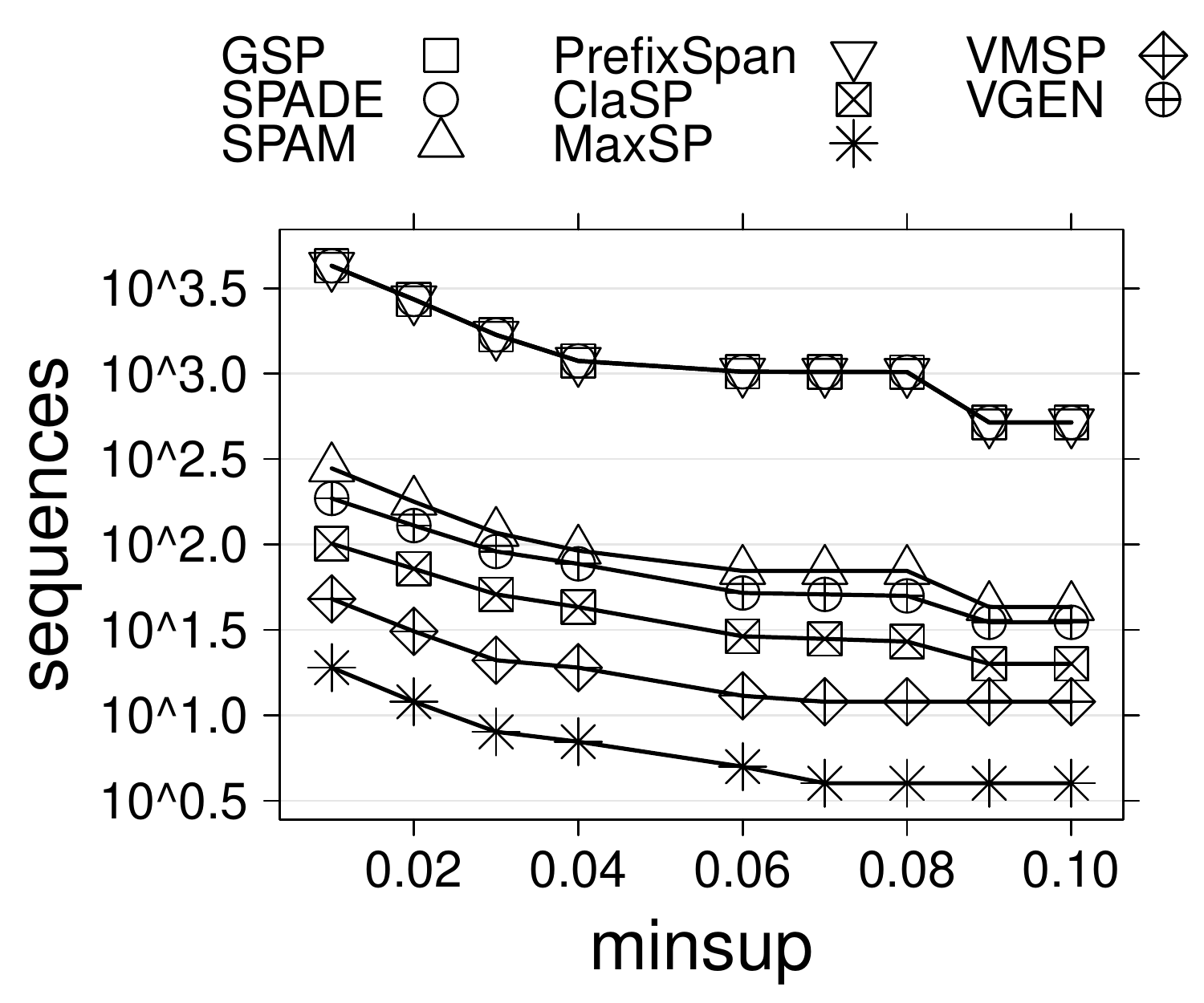}
	\end{subfigure}
	\caption{Time, memory and number of sequences for different minimum support values}
	\label{fig:app-time-memory}
\end{figure*}

For our problem, we compared the performance of the following well-known algorithms. GSP, Spade, and Spam (Apriori-all BFS and DFS); PrefixSpan (Pattern-growth DFS); ClaSP (Closed); MaxSP, and VMSP (Maximal); VGEN (Generator). Figure~\ref{fig:app-time-memory} shows the time and memory used, as well as the number of sequences generated, for the considered algorithms across different minimum support values. We can see that VMSP is very competitive in terms of time and memory used. PrefixSpan, which explores all sequential patterns, exhibits similar time and memory usage. However, PrefixSpan generates roughly two orders of magnitude more sequences that VMSP. This means that \name would have to store a larger amount of metadata than it would with VMSP. MaxSP generates the fewer sequences, but it is not efficient in terms of memory usage.

\spara{Our choice.} For our specific problem, we want to discover all maximal sequential patterns, since we want to prefetch the largest number of items as possible (from large sequences) and trigger the prefetching process in the first item. For example, having the frequent sequence $S = \langle a, b, c, d, e\rangle$, we want to prefetch $\{b, c, d, e\}$ when the application requests $a$, which is redundant with prefetching ${c, d, e}$ when the user requests $b$ if we have a frequent sequence $S' = \langle b, c, d, e \rangle$ with the same support as $S$.

We decided to adopt the VMSP as our default algorithm for discovering maximal sequential patterns, since it is a modern fast and open-source algorithm, offering parameters to constraint the pattern length and gap. The former is useful because we are only interested in sequences with length between 3 and 15 (reasonable value for most applications); and the latter is important because we only want to have consecutive itemsets of a pattern that appear consecutively in a sequence (i.e., no gap is allowed).


\spara{Data post-processing.} We limit the number of sequences that we maintain in memory in order to bound its usage by metadata. In some cases, the number of discovered sequences may be greater than this threshold, and thus we have to select a subset of those sequences. To select them, we create a ranking where we multiply the length of the sequence by its support; i.e., the larger sequences we have, and the higher is their support, the better. After, we only select the top ones, according to the ranking, that can fit in the metastore.

\section{\name Design and Implementation}
\label{sect:architecture}
In this section we describe the general architecture, design choices, and implementation of our system, \name, including how pattern discovery is done, how discovered patterns are used to leverage prefetching, and how our cache operates with minimum pollution.
%
\begin{figure}
	\centering
	\includegraphics[width=0.7\columnwidth]{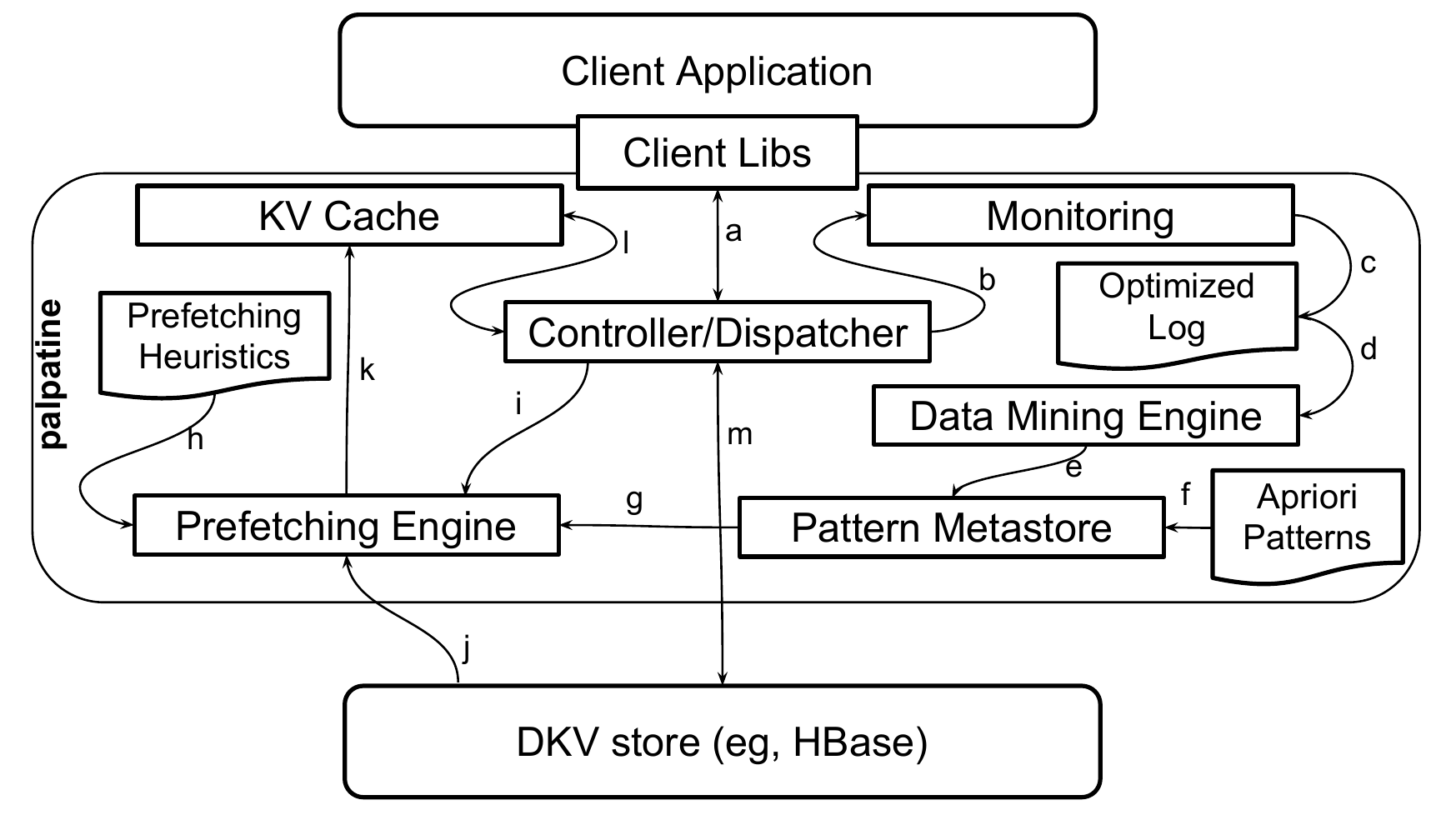}
	\caption{Architecture overview}
	\label{fig:architecture}
\end{figure}

\subsection{Architecture and work flow}

Figure~\ref{fig:architecture} depicts the architecture overview of \name, which is integrated with the client libraries of the DKV store (i.e., at the application level). The general work flow is described as follows. Client applications issue read/write requests through the DKV store client libraries and APIs. These requests are intercepted by the Controller component (step \emph{a}), which dispatches them to the Monitoring component (step \emph{b}). The Monitoring analyzes the requests and appends all relevant information to a backlog whose structure is optimized for running Data Mining algorithms (step \emph{c}). After a sufficient number of requests have been logged, the respective data (typically a file) is sent to the Data Mining Engine (step \emph{d}). In turn, this engine runs algorithms to assess the most frequent access sequences. The discovered sequences along with any apriori known sequences are stored in the Pattern Metastore component (steps \emph{e} and \emph{f}).

The Prefetching Engine component takes the frequent sequences from the metastore (step \emph{g}), and along with Prefetching Heuristics (step \emph{h}) and requested data (step \emph{i}), decides what data pieces to prefetch from the DKV store and fetches them asynchronously in the background (step \emph{j}). The fetched data pieces are then stored in the KV Cache (step \emph{k}). When new read requests are issued (step \emph{a}), the data can be served either from the cache (step \emph{l}) or slower from the DKV store (step \emph{m}).

\spara{Handling reads.} Upon a client read request, the \emph{Controller} checks if the requested data containers are present in cache. If all data containers are living in cache, their corresponding values are returned to the client and no additional action is taken. Otherwise, containers that are not living in cache are: $(i)$ retrieved from the data store; $(ii)$ aggregated with cached data containers if needed; $(iii)$ returned to the client; and $(iv)$ cached to serve possible further requests. Simultaneously in the background, the prefetching process is initiated. The \emph{Controller} queries the \emph{Pattern Metastore} for possible sequences that start with the requested data containers. If there are some sequences, \emph{Prefetching Heuristics} determines which data containers comprising them are fetched from the data store and cached in memory, since their access is likely to happen in subsequent requests. These subsequent data containers in a sequence are prefetched progressively and in order so that they are cached before they are requested (i.e., fetching complete sequences at once could introduce major delays).

\spara{Handling writes.} When a write request is issued, the associated data container is updated both in cache and in the data store.

\subsection{Pattern Discovery}
\label{sect:pattern-discovery}

Frequent patterns of sequentially accessed items to the DKV store by a user or application may either remain roughly static or change over time. If patterns do not change significantly, we only need to perform the mining process a single time to populate the pattern metastore (this is the most efficient). Otherwise, we have an online mechanism where we are continuously monitoring data accesses and performing data mining on new log files to discover fresh patterns. In this case, the frequency of the mining process can be set based on time (e.g., every \emph{x} minutes/hours interval) or based on the size of the log file, which contains data since the last mining execution. As this process can yield some overhead in terms of CPU and memory, we always perform it in the background with the lowest possible priority (cf. Sect.~\ref{sect:implementation-details}).

The minimum support value (cf. Sect~\ref{sect:mining-algos}) of our adopted algorithm, VMSP, which specifies the percentage of times a sequence appears in the data store log, is dynamically set: we start with a high value (for our problem, 50\%) and, if the number of resulting frequent sequences is not satisfactory, we keep decreasing the minimum support until we get enough patterns (the minimum support value range, decrease step, and minimum number of frequent patterns can be manually defined). The lesser this value is, the more are the sequences we get as a result, and the more is the likelihood of performing non useful prefetching. If we end up with many sequences, however, we only take the most representative ones in the next step (cf. Sect.~\ref{sect:mining-algos}).

\begin{figure}[h!]
	\centering
	\includegraphics[width=0.5\columnwidth]{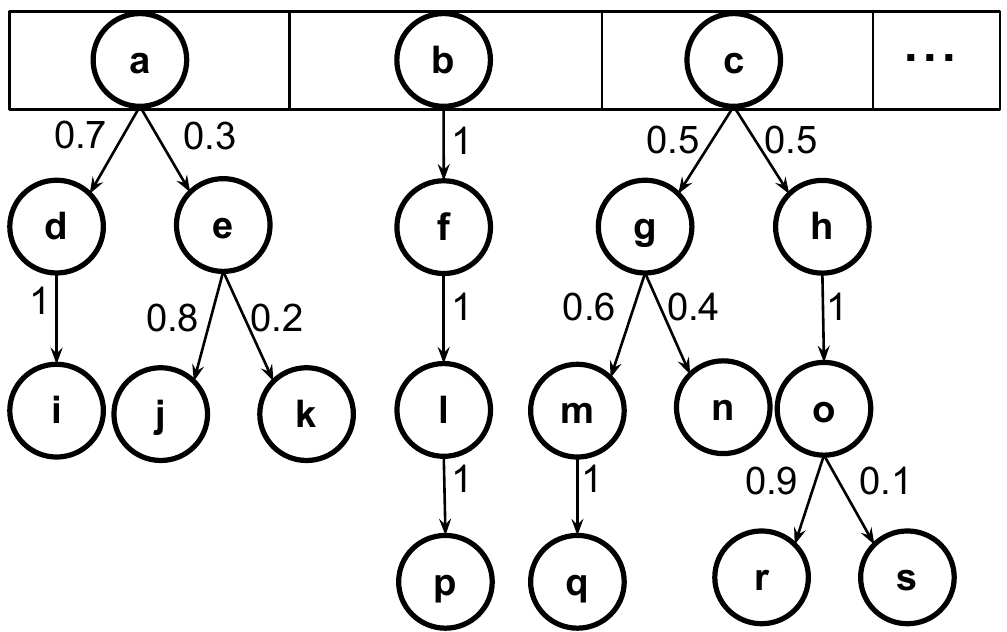}
	\caption{Probabilistic model}
	\label{fig:sequences}
\end{figure}

After having the pattern metastore populated with the most frequent sequences of accessed items, we build probabilistic trees (akin to Markov Chains). Figure~\ref{fig:sequences} depicts this data structure, where the nodes are the accessed items and branches represent the dependencies between items with a given probability. In this example, we have 8 sequences starting by the items $\{a, b, c\}$. In the first tree, starting by item $a$, we have the sequences $\langle a, d, i\rangle$, $\langle a, e, j \rangle$, and $\langle a, e, k \rangle$. Once item $a$ is accessed there is a 70\% probability of going to item $d$ and a 30\% probability of going to item $e$; and after subsequence $\langle a, e \rangle$ has been accessed, either item $j$ or $k$ can be accessed with probabilities of 80 and 20\% respectively. These probabilities are calculated based on the frequencies of the observed sequences, and they are used at runtime to make prefetching decisions according to cache parameters (e.g., cache size). 

\subsection{Prefetching Heuristics}

The prefetching needs to be timely, useful and with low overhead. Based on the stored probabilistic trees, \name can employ different strategies to determine what data should be fetched from the DKV store and loaded into cache. These strategies range from conservative to more progressive, according to runtime cache parameters (i.e., the more available resources, the more items can be prefetched).

\begin{figure}[h!]
	\centering
	\includegraphics[width=0.5\columnwidth]{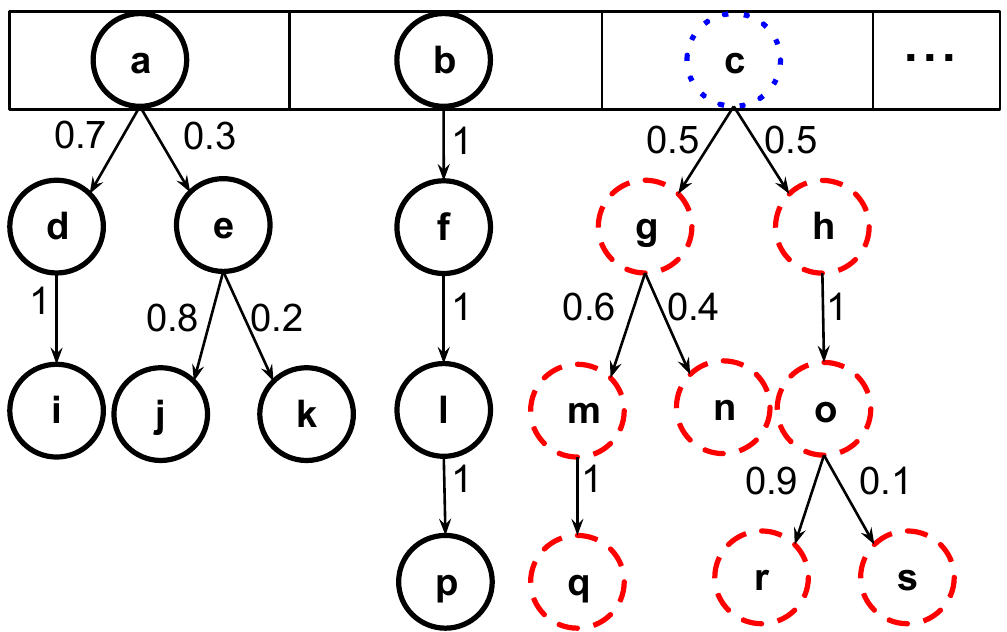}
	\caption{Fetch all: item \emph{c} is requested and all items from its tree are fetched.}
	\label{fig:heuristic-1}
\end{figure}

Each item requested by the client application is matched against an index containing all root nodes of the trees. If the requested item corresponds to a root node, a prefetching context is created. This context maintains state that varies across different heuristics. Note that multiple contexts may run in parallel, since each and every single request can potentially create a new context. Following, we describe the heuristics supported by our system:

\spara{Fetch all.} When an item matching a root node of a stored tree is requested, the entire tree originated in that node is prefetched from the DKV store into cache. This heuristic is exemplified in Figure~\ref{fig:heuristic-1}. It has the best accuracy, since it prefetches more items, but also greatest pollution potential.

\begin{figure}[h!]
	\centering
	\includegraphics[width=0.5\columnwidth]{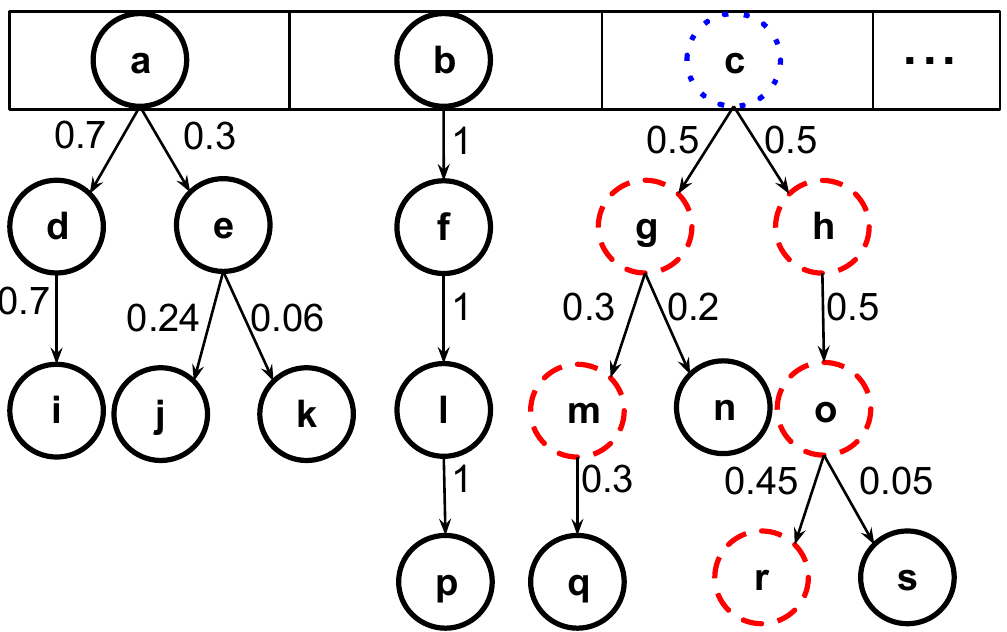}
	\caption{Fetch the most probable 5 items: item \emph{c} is requested and the 5 items with higher cumulative probability are fetched.}
	\label{fig:heuristic-2}
\end{figure}

\spara{Fetch the most probable \emph{n} items.} When an item matching a root node of a stored tree is requested, the top \emph{n} items with higher cumulative probability are prefetched in level-order. The cumulative probability of a node is the probability of an item being requested when starting from the root and not from its parent node. Moreover, the parameter \emph{n} can be manually defined and balanced to offer a good compromise between accuracy and pollution. 
Figure~\ref{fig:heuristic-2} exemplifies this heuristic for $n=5$.

\begin{figure}[h!]
	\centering
	\includegraphics[width=0.2\columnwidth]{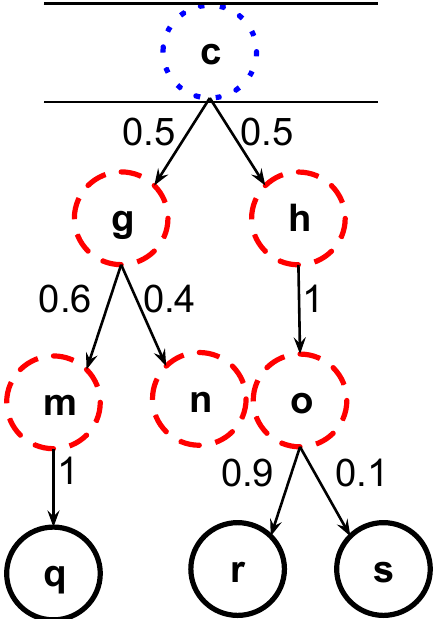}~~~~~
	\includegraphics[width=0.2\columnwidth]{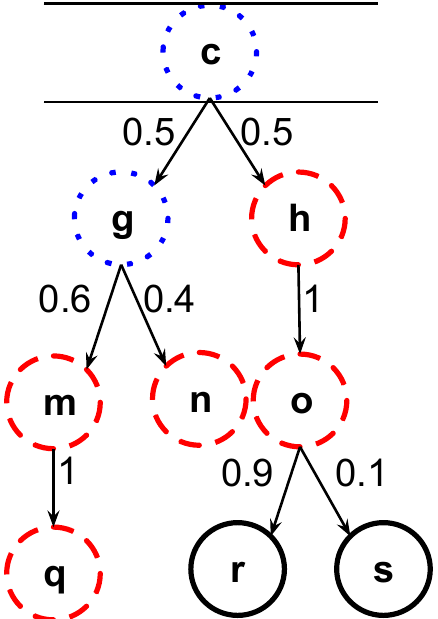}
	\caption{Fetch progressively with hints: item \emph{c} is requested and the next two levels of the tree are fetched (left side). Then, item \emph{g} is requested and item \emph{q} is fetched (right side).}
	\label{fig:heuristic-3}
\end{figure}

\spara{Fetch progressively with hints.} Firstly, when an item matching a root node of a stored tree is requested, the items of the next \emph{n} levels of the tree are prefetched. Secondly, this heuristic checks whether subsequent requests correspond to a subsequence without gaps from the root. If not, it means that these requests do not comprise a frequent sequence and, therefore, no further action is taken. Otherwise, for each subsequent matched request, we prefetch the items of the next non cached level that can be reached by the actual requested subsequence (i.e., we cut the tree as requests are being made) until we reach the maximum depth of the respective tree. Figure~\ref{fig:heuristic-3} exemplifies this heuristic for $n = 2$. Moreover, the parameter \emph{n}, which can be manually configured, should be small enough to avoid fetching many unnecessary items (default $n=2$). This heuristic is the best in terms of maintaining a good balance between accuracy and pollution.

\subsection{Cache}

Our cache comprises two distinct spaces: one space where all actual requested items go (main space), and a space where all prefetched items go (preemptive space). The size of the main space can be manually configured; and a percentage of that size is allocated to the preemptive space (10\% by default). This separation of spaces avoids cache pollution and parallelizes the potential churn caused by prefetching unnecessary items. Further, each item requested by the application is stored in the main space, either by normally fetching it from the back store (slower) or by moving it from the preemptive to the main space (in case the item was prefetched before). 

For each of the two spaces, we maintain an independent cache, where we rely on LRU as default replacement policy. This independence allows for fast parallel access in both spaces.

Regarding cache coherence, we invalidate items in both cache spaces when new values for those items are written by the client application. In fact, we replace the old values by the new values directly in cache, considering the respective items as the most recent ones, while also performing the write operation to the back store asynchronously. In addition, we have a base mechanism to maintain caches coherent across multiple clients. This mechanism is based on a monitoring module (e.g., co-processors in HBase or triggers in Cassandra) at the data store level that simply notifies caches of updated items, so that they can be invalidated in case they are cached. Note that writing conflicts are handled by the underlying data store, typically by following the last-write-wins strategy.

\subsection{Implementation Details}
\label{sect:implementation-details}

Palpatine is integrated with the HBase client libraries (HBase used version: $1.2.4$). The source code of its prototype is publicly available here~\cite{code}. For the sequential pattern mining algorithms, we resorted to the SPMF~\cite{fournier2016spmf} (version $2.17$), an open-source data mining library. 

To make use of our system, applications only need to use our provided HBase client libraries instead of the original libraries, which maintain the exact same API. As such, applications that use HBase as back store do not need to be modified (i.e., \name is transparent to applications).

We rely on a thread with low priority for performing the data mining asynchronously in the background. We make use of the VMSP algorithm implementation in the SPMF library. Its input comprises: $(i)$ a log file, representing a sequence database; $(ii)$ the minimum support threshold (a value in [0,1] representing a percentage); and $(iii)$ the maximum sequential pattern length in terms of items.

We represent the probabilistic trees as hash tables of trees whose keys represent the first items of the frequent sequences. These items correspond to data containers, which can be a table, row, column (including family and qualifier), or any combination of these. For example, a path of a tree may represent a sequence of table and columns that are accessed for a given row. We make prefetching requests to the backstore first level-order, and second probability-wise, through a tree iterator, so that the subsequent items in the sequence requested by the application are the first to be cached. Each heuristic has a different implementation of this iterator. Further, we batch prefetching requests as much as possible on a per table basis so that multiple requests can be performed in very few round trips. First level trees however are not batched so that we can anticipate subsequent application requests in a timely fashion.

\section{Evaluation}
\label{sect:evaluation}
To validate and demonstrate the effectiveness of \name, we conducted an experimental evaluation of our prototype. The main objective of this evaluation was to assess: $(i)$ the performance of the data mining process (Section~\ref{sect:mining}); $(ii)$ the benefits of \name in terms of accuracy, latency, and throughput (Section \ref{sect:gains}); and $(iii)$ our system overhead in terms of runtime (Section~\ref{sect:overhead}).

\spara{Workloads.} To evaluate \name, we resorted to two realistic benchmarks: \emph{SEQB}, a benchmark we developed to stress our system and exercise commonly observed types of patterns; and \emph{TPC-C}, the well known On-Line Transaction Processing Benchmark that simulates a wholesale supplier. The source codes of both benchmarks are publicly available~\cite{benchmark,tpcc}.

\emph{SEQB} is a benchmark that can generate a myriad of different types of access patterns that follow empirical distributions (i.e., resulting from real world observations). It allows us to exercise different parameters of our system while controlling the frequency of recurrent patterns.

SEQB comprises two stages. In the first stage, it starts by populating the back store with blocks of random bytes. Then, the workload starts running and read/write operations are issued to the data store according to a zipfian distribution (to simulate realistic access patterns we make some data containers to be accessed more often than others). At this stage, Pattern Metastore is empty and, as such, no prefetching is performed. When the workload completes, after a specified number of operations, we end this stage by furnishing the Pattern Metastore (otherwise empty) with frequent observed access sequences (as described in Section~\ref{sect:architecture}). In the second stage, our benchmark starts running the same workload and observed patterns with prefetching fully operational. 

\emph{TPC-C} simulates a complete environment where users execute transactions on a data store. It focuses on the the main transactions performed by order management systems, representing thus any industry that must manage, sell, or distribute a product or service. These transactions include entering and delivering orders, recording payments, checking the status of orders, and monitoring the stock levels at the warehouses. The frequency of the individual transactions are modeled after realistic scenarios~\cite{tpcc-detail}. 
Further, TPC-C provides a set of interesting characteristics to our problem, such as: significant disk input/output; non-uniform distribution of data access through primary and secondary keys; databases consisting of many tables with a wide variety of sizes, attributes, and relationships; and contention on data access and update.

Similarly to SEQB, we considered two execution stages for TPC-C. In the first stage, we observe and capture varying access patterns to the data store, with prefetching deactivated. As for the second stage, the workload is executed with prefetching activated, corresponding to our steady-state. Apart from the data mining process, our evaluation is based on this steady state.

\spara{Setup.} \name experimental setup and parameters are specified as follows.

\noindent Pattern mining:
\begin{squishlist}
    \item Minimum and maximum sequence length: 3 and 15
    \item Maximum gap: 1 (i.e., no gap)
    \item Minimum support: from 0.01 to 0.1
    \item Pattern Metastore capacity: $10,000$ sequences of up to 15 elements  
\end{squishlist}

\noindent Cache and prefetching:
\begin{squishlist}
    \item Cache size: 2, 4, 8, 16, 32, 64, 128, 256 MB
    \item Heuristics used: fetch-all, fetch-top-n (top 5 most probable items), fetch-progressively
\end{squishlist}    

\noindent SEQB benchmark:
\begin{squishlist}
  \item Data containers: $2,300,000$ blocks of 1000 random bytes
  \item Access sequences minimum and maximum size: 3 and 10
  \item Number of frequent sequences bias: between 80 and $10,240$
  \item Zipfian exponent: 0.5, 1.0, 1.5, 2.0, 2.5, 3 (higher values lead to higher incidence of operation requests over frequently accessed sequences)
  \item Total number of sessions of operations: $10,000$
  \item Type of workload: read intensive
\end{squishlist}

\noindent TPC-C benchmark:
\begin{squishlist}
	\item Scale parameters: 10 districts/warehouses, 3000 customers per district, 900 initial orders per district, 100,000 items
	\item Data containers: blocks of a maximum of 500 bytes
	\item Total number of sessions of operations (or transactions) in the second execution stage: 350
	\item Sequence factor: 0.1, 0.2, ..., 2 (represents the percentage of second stage number of sessions/transactions that are used in the first stage for the data mining process)
\end{squishlist}

We compared \name with a baseline, corresponding to the standard version of HBase and respective client libraries without any modifications.

\spara{Setting.}
All tests were conducted using two machines with an Intel Core i7-2600K CPU at 3.40GHz, 11926MB of available RAM memory, and HDD 7200RPM SATA 6Gb/s 32MB cache, connected by a 100Mbps network. One machine was used to run the HBase DKV store and the other to run our benchmarks as client applications.

We chose this setting to analyze the behavior of the system on a per-node basis, as worst case, since our relative gains are further amplified when using a cluster of machines. The running environment consisted of Ubuntu 16.04.3 LTS (GNU/Linux 4.4.0-97-generic x86\_64), Java OpenJDK 1.8.0\_151, HBase 1.2.4 and SPMF 2.17.


\subsection{Data mining}
\label{sect:mining}

The discovery of sequential patterns, and furnishing of the Pattern Metastore, occurs in the first stage of the workload. Our adopted sequential pattern mining algorithm, VMSP, discovers only frequent \emph{maximal} sequential patterns and includes parameters such as the minimum/maximum pattern length and maximum gap (cf.~\ref{sect:mining-algos}), which allow to improve the algorithm performance, and reduce the number of useless sequences, to a great extent. The minimum support (minsup) parameter is the only one that we vary across executions.


\begin{figure}
	\centering
	\begin{subfigure}[t]{0.49\columnwidth}
	  \includegraphics[width=\columnwidth]{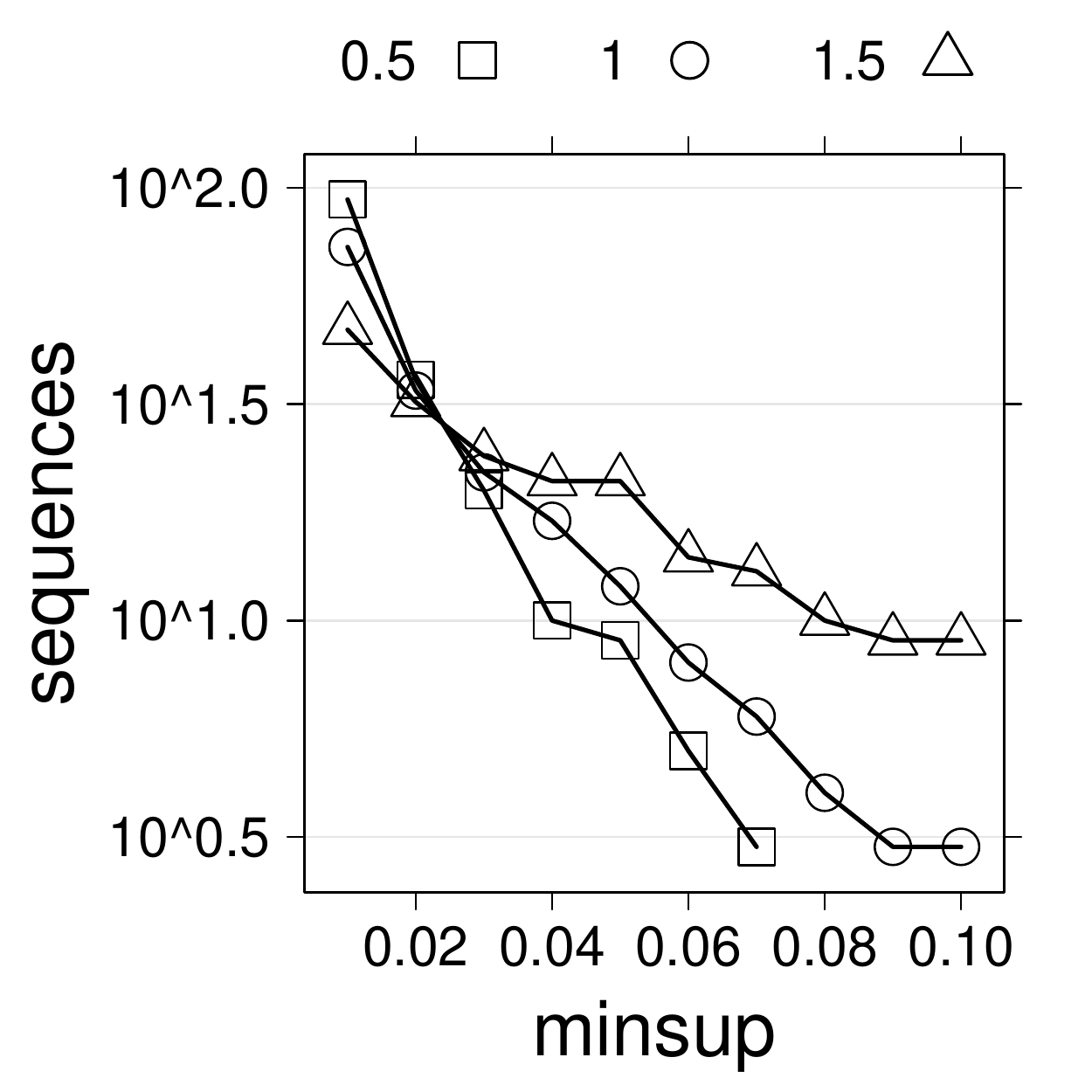}
	  \caption{SEQB, exp=\{0.5,1,1.5\}}
	  \label{fig:seqb-number-sequences}
	\end{subfigure}
	\begin{subfigure}[t]{0.49\columnwidth}
  	  \includegraphics[width=\columnwidth]{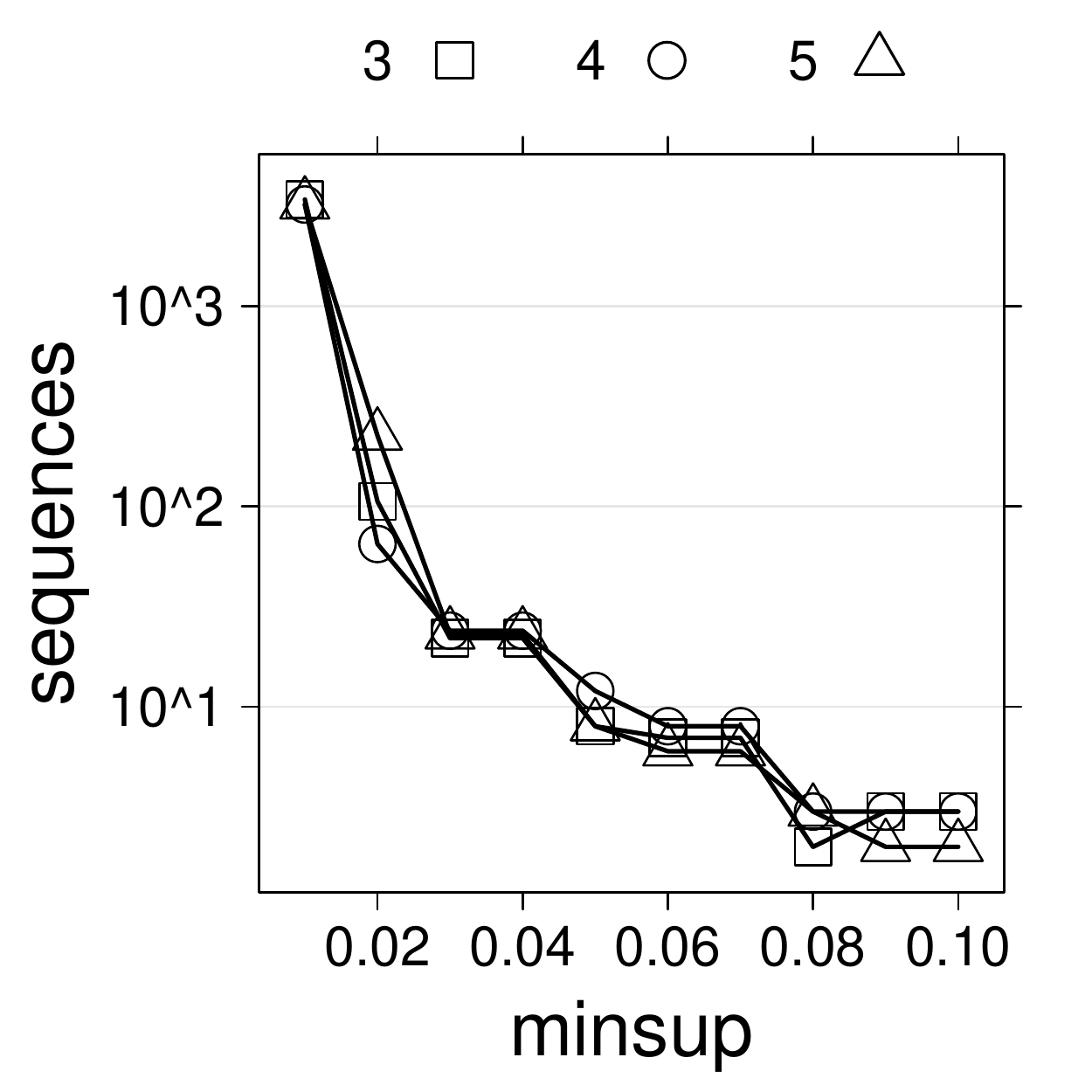}
  	  \caption{TPC-C, sequence factor=\{3,4,5\}}
  	  \label{fig:tpcc-number-sequences}
  	\end{subfigure}
	\caption{Number of sequences for different minsup values}
	\label{fig:number-sequences}
\end{figure}

Figure~\ref{fig:number-sequences} shows the number of sequences, in logarithmic scale, obtained for SEQB and TPC-C while varying the minsup. The low values of minsup ensure that we are getting as many sequences as possible. For SEQB, Figure~\ref{fig:seqb-number-sequences}, we may first observe that the number of sequences decreases abruptly as the minsup increases, and the number of sequences around a minsup of $0.1$ is not exceedingly large for the Pattern Metastore.

Second, we can see that larger exponents, that increase the recurrence of frequent patterns, lead mostly to higher number of sequences. This behavior is expected, since increasing the pattern recurrence also increases their support within the data store. However, larger exponents reduce the expressivity of other not so frequent patterns during the workload execution, and that is the reason why these exponents initially yield lower number of sequences for a minsup lower than $0.02$.

In Figure~\ref{fig:tpcc-number-sequences}, TPC-C exhibits a smoother decrease in the number of sequences as the minsup increases. This is explained due to the non-uniform distribution of data accesses in this benchmark; i.e., within the set of accessed sequences, there is no subset that stands out to a very high degree in relation to any other. Most frequent sequences were concentrated with a minsup lower than $0.03$, and the maximum number of sequences observed, with $minsup=0.01$, does not exceed the Pattern Metastore capacity.

The time and memory spent on this mining process are naturally higher for lower values of minsup (or higher number of sequences discovered). In the worst case ($minsup=0.01$), the maximum time and memory we observed is given in the table below.

\begin{center}
\begin{tabular}{ |c|r|r| } 
 \hline
 benchmark & max time (ms) & max memory (MB)\\
 \hline
 SEQB & 452 & 67\\ 
 TPC-C & 3152 & 1236\\ 
 \hline
\end{tabular}
\end{center}

In terms of time, the process takes less than 5 seconds for each benchmark, using traces with a size corresponding to that of a complete execution of the workloads. As for memory, SEQB used less than 100MB and TPC-C around 1.2GB. Note that in an online process, we would perform the mining on smaller chunks of traces, which would lead to faster and less memory intensive executions. We consider the values we obtained do not pose a major impact to system resources. Moreover, these values are fairly acceptable, since we can support up to 720 data mining processes of up to 5 seconds and 1.5GB in a dedicated \emph{t2.small} Amazon EC2 VM with a cost of 0.023USD/hour~\cite{amazonec2}.





\subsection{Gains}
\label{sect:gains}

The benefits of our system come from having an effective prefetching mechanism. We start by evaluating the accuracy (or precision) of \name, and then we evaluate the impact it has on system latency, throughput and runtime. Finally, we assess the overhead of \name.

\spara{Accuracy.} The hit rate ($ cacheHits / numberOfAccesses $) provides a global metric of our cache, which includes all accesses that were served by our cache. The precision captures the part of the hit-rate that comes solely by immediate success with the prefetching. Later hits to prefetched cache items, while also attributable in part to prefetching, will be driven by the LRU behavior of the cache. The precision is given by: 

\fbox{$ precision = prefetchHits / numberOfPrefetches $}

A prefetch hit occurs when a prefetched item is accessed for the first time while its value is still present in cache. Subsequent accesses to that same item do not account as prefetch hits, but rather as cache hits. For example, consider the tree of sequences starting by \emph{a} in Figure~\ref{fig:sequences}: the application requests \emph{a}, and, with the heuristic \emph{fetch-all} (for instance), we prefetch items $\{d, e, i, j, k\}$; if it is a frequent sequence (e.g., $\langle a,d,i \rangle$, $\langle a,e,j \rangle$) the maximum precision we can have with this sequence for this tree is $2 / 5$; to achieve a precision of 100\%, the other non accessed prefetched items need to be requested by other sequences.


\begin{figure*}
	\centering
  	\begin{subfigure}[t]{0.24\textwidth}
	  \includegraphics[width=\columnwidth]{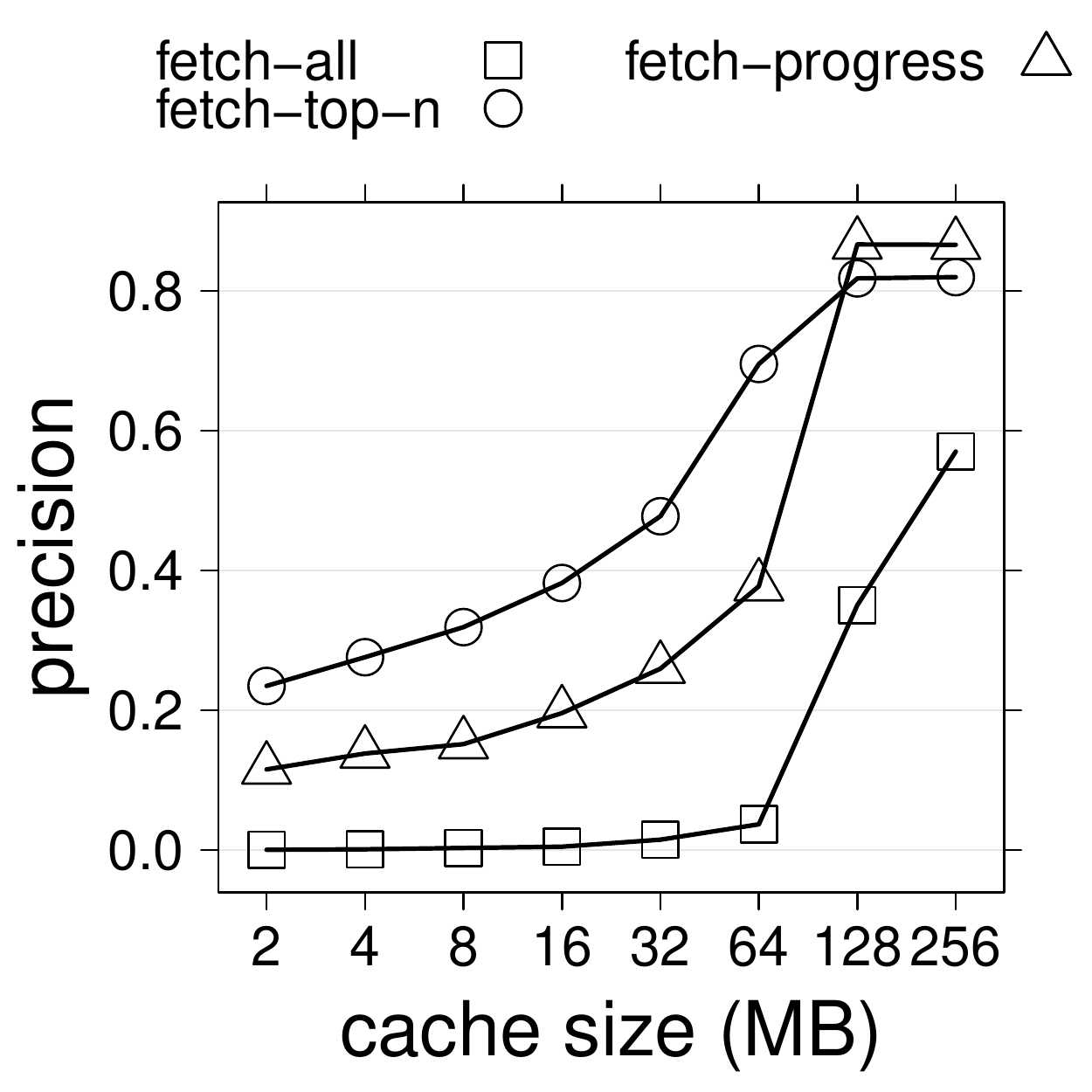}
	  \caption{Precision}
	  \label{fig:precision-cache-size}
	\end{subfigure}
	\begin{subfigure}[t]{0.24\textwidth}
  	  \includegraphics[width=\columnwidth]{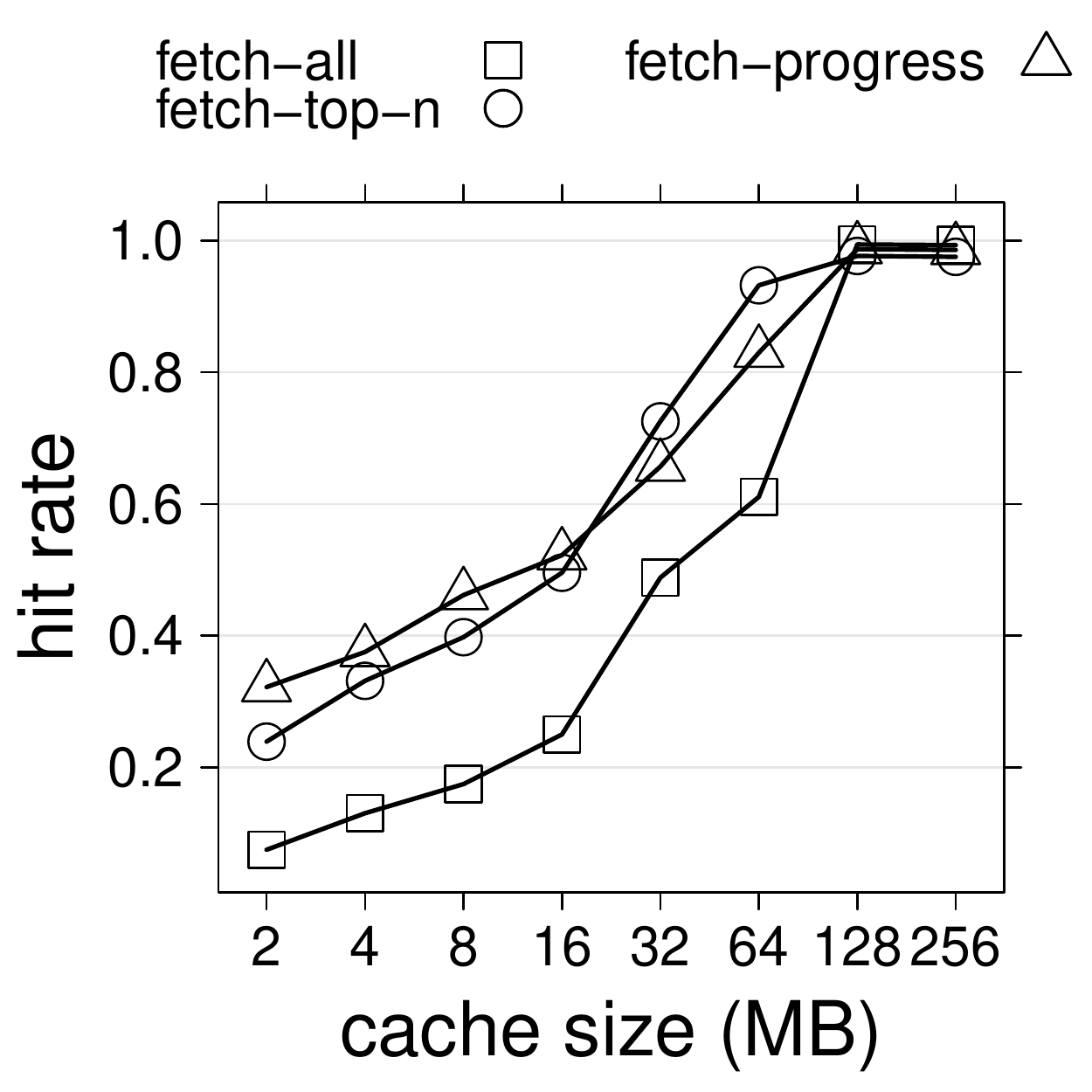}
  	  \caption{Hit rate}
  	  \label{fig:hit-rate-cache-size}
  	\end{subfigure} 
	\begin{subfigure}[t]{0.24\textwidth}
	  \includegraphics[width=\columnwidth]{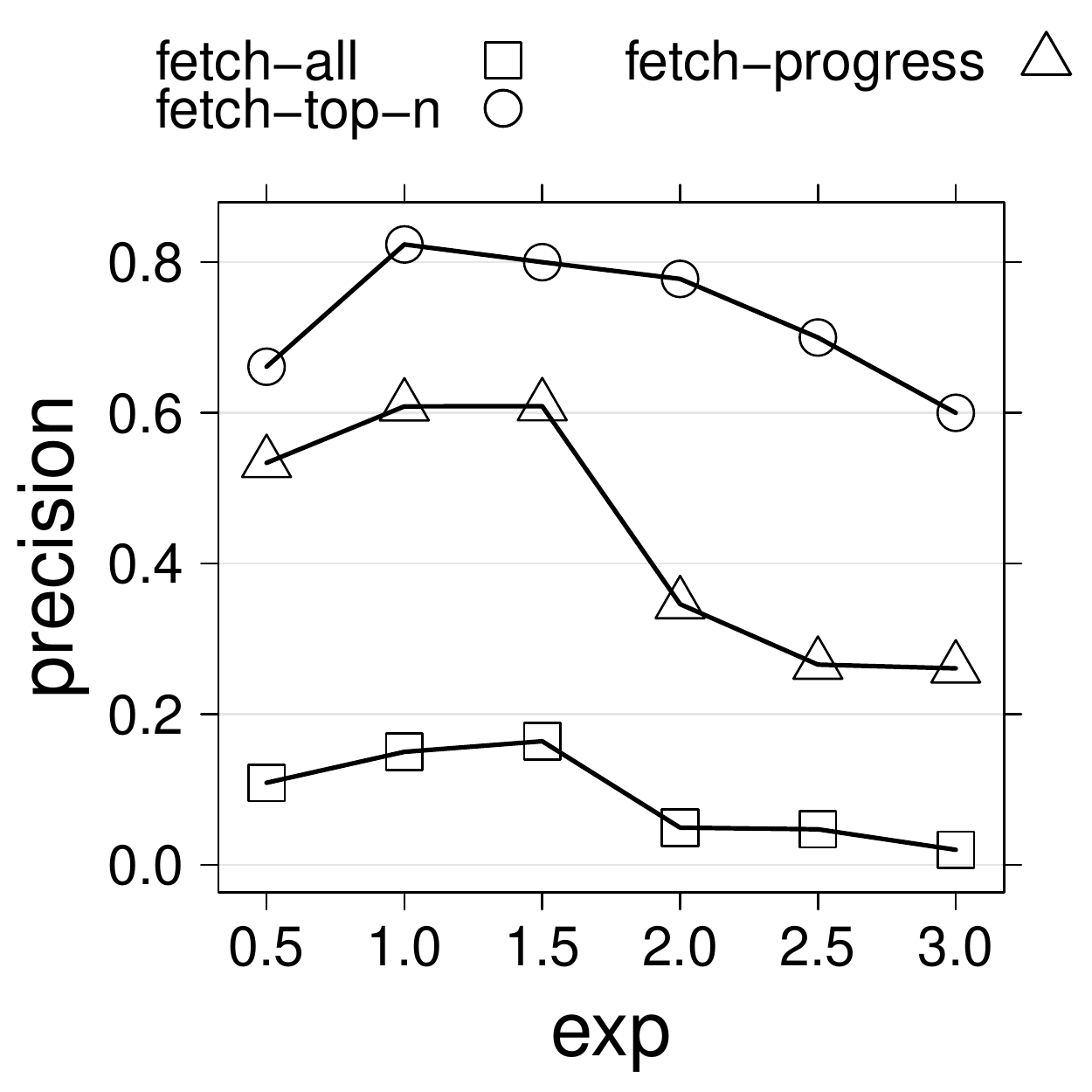}
	  \caption{Precision}
	  \label{fig:precision}
	\end{subfigure}
	\begin{subfigure}[t]{0.24\textwidth}
  	  \includegraphics[width=\columnwidth]{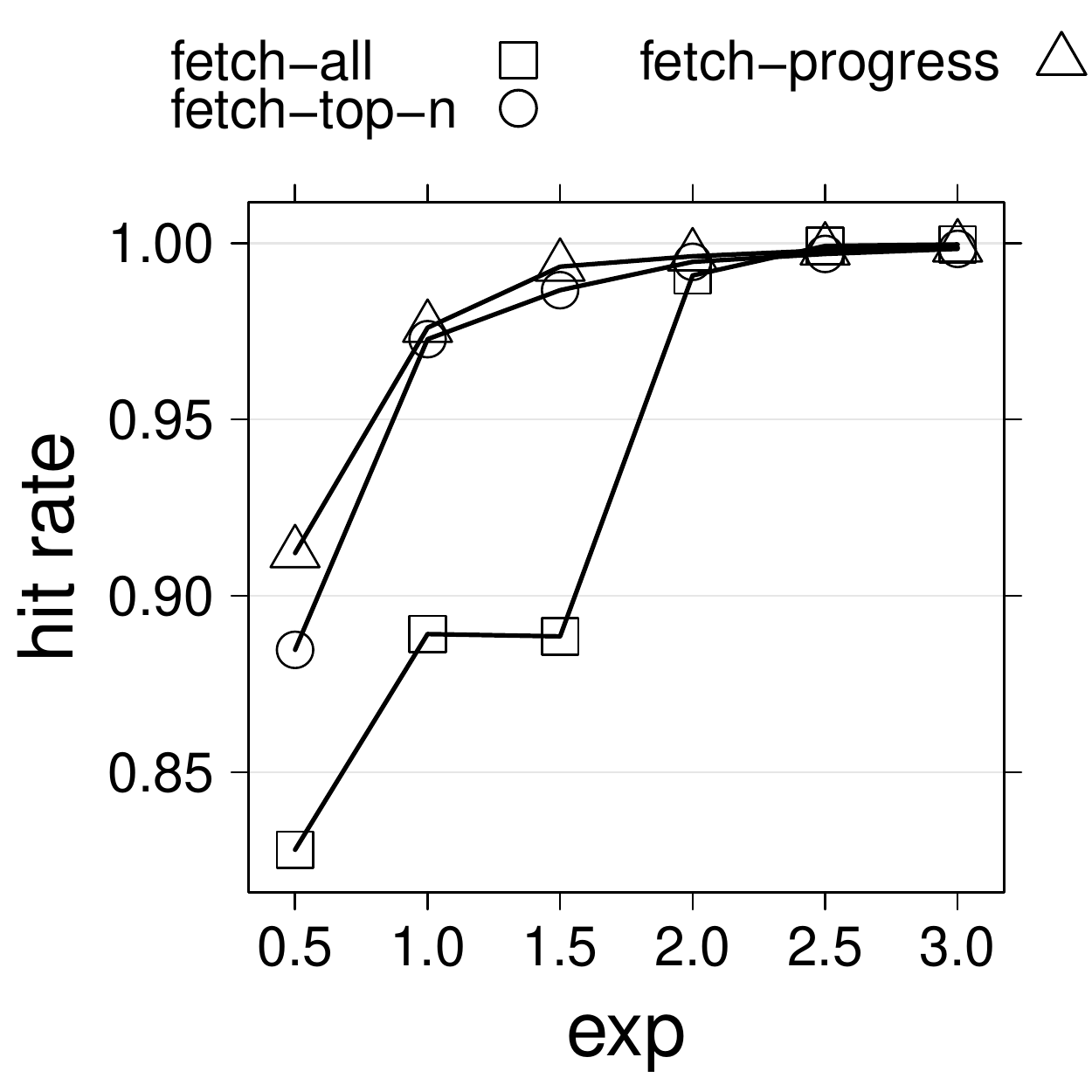}
  	  \caption{Hit rate}
  	  \label{fig:hit-rate}
  	\end{subfigure}
  	
	\caption{SEQB prefetch precision and cache hit rate}
	\label{fig:seqb-precision-hit-rate}  	
 \end{figure*}
 
For SEQB, figures~\ref{fig:precision-cache-size} and \ref{fig:hit-rate-cache-size} show the impact of the cache size on precision and hit rate for different heuristics (as specified in Section~\ref{sect:pattern-discovery}). For the zipfian distribution we used an exponent of $1.0$ as the default value. We can observe that the precision and hit rate increase with the cache size, which is expected since caching more elements leads to higher number of prefetch and cache hits. With a large cache size of $128-256MB$ we are able to reach almost 100\% of cache hits, where about 60-80\% of the prefetches resulted in hits (i.e., high proportion of useful prefetching and low cache pollution). The heuristic \emph{fetch-all} yielded the lowest precision and hit rate, indicating that the proportion of useless prefetches (i.e., prefetches that were not hit) was higher than in other heuristics (around 20\%). With higher precision, we have the \emph{fetch-top-n} heuristic. The reason for such result is that this heuristic is the one that performs the fewest number of prefetches. In between, we have \emph{fetch-progress} which provides the best balance between the number of performed prefetches and prefetch hits.
 
For the remaining of the experiments with SEQB, we decided to set the cache size to 32MB since it already provides relevant improvements in hit rate (with the default zipfian exponent of 1.0) and limits cache memory overhead to an acceptable value for any application runtime however constrained (e.g., running as microservice in a Docker container, or in dedicated cache instances placed closer to the network edge, or even running locally in a mobile device alongside the front-end of the application).

Figure~\ref{fig:precision} shows the precision of SEQB for different heuristics and exponents of the zipfian distribution, which specify the recurrence of accessed frequent sequences during the workload execution. We can observe that the precision drops smoothly across different zipf exponents for each heuristic. The reason for this behavior is that the number of prefetched elements increases more than the number of the corresponding hits, as we increase the exponent. In other words, since there is a higher bias towards fewer sequences, given by larger zipfian exponents, we get more prefetches yet less prefetch hits (note that, unlike a cache hit, a prefetch hit can just happen 1 time for a given element).

Figure~\ref{fig:hit-rate} depicts the overall hit rate of the \name cache for different exponents and heuristics. We may observe that as the recurrence of frequent patterns (exp) increases, the hit rate also increases, since with less entropy we can anticipate more requests. Apart from \emph{fetch-all}, all heuristics behave similarly, going from an hit rate of around 80\% to almost 100\%, meaning that the majority of the requests were satisfied by our cache.

 \begin{figure*}
  	\begin{subfigure}[t]{0.24\textwidth}
	  \includegraphics[width=\columnwidth]{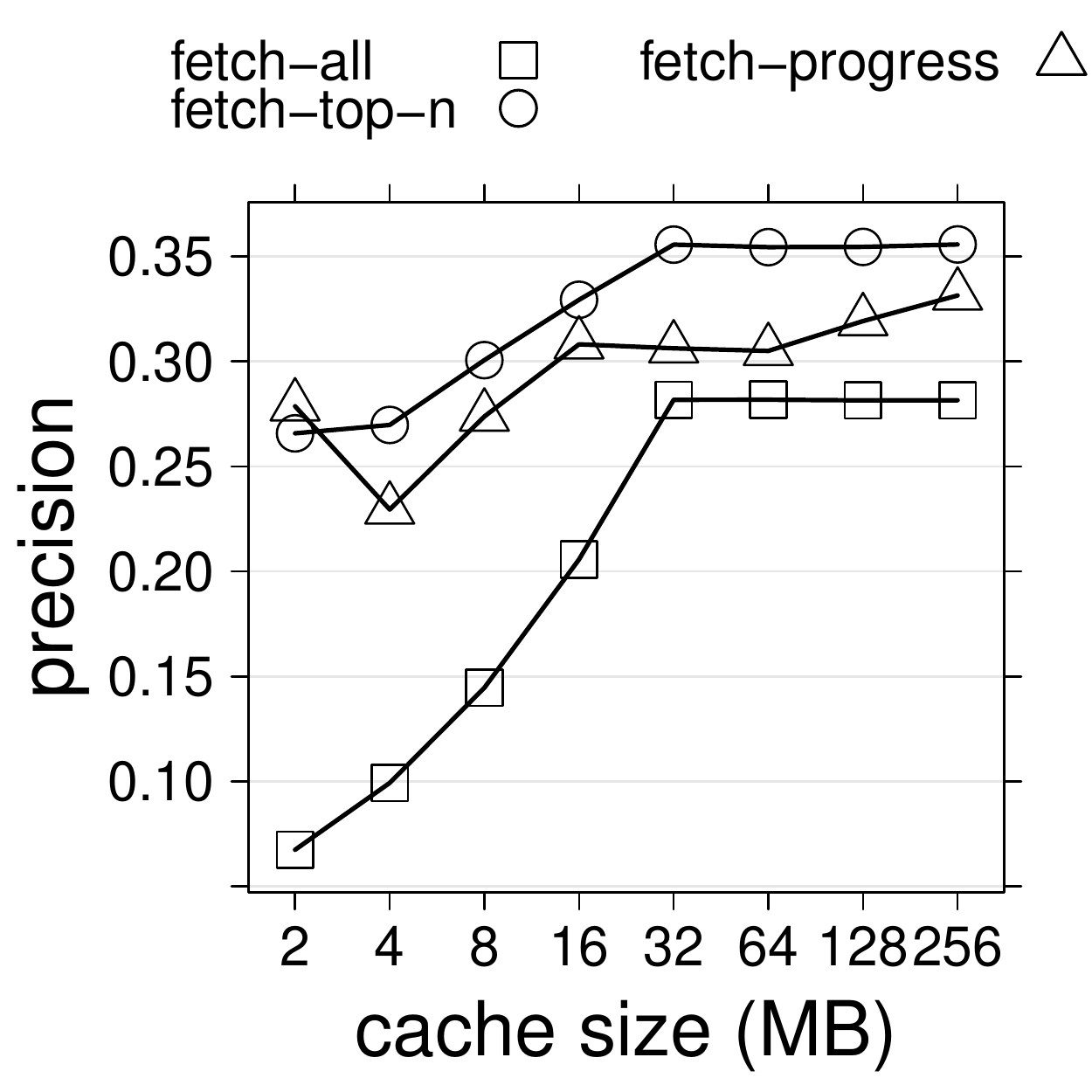}
	  \caption{Precision}
	  \label{fig:tpcc-precision-cache-size}
	\end{subfigure}
	\begin{subfigure}[t]{0.24\textwidth}
  	  \includegraphics[width=\columnwidth]{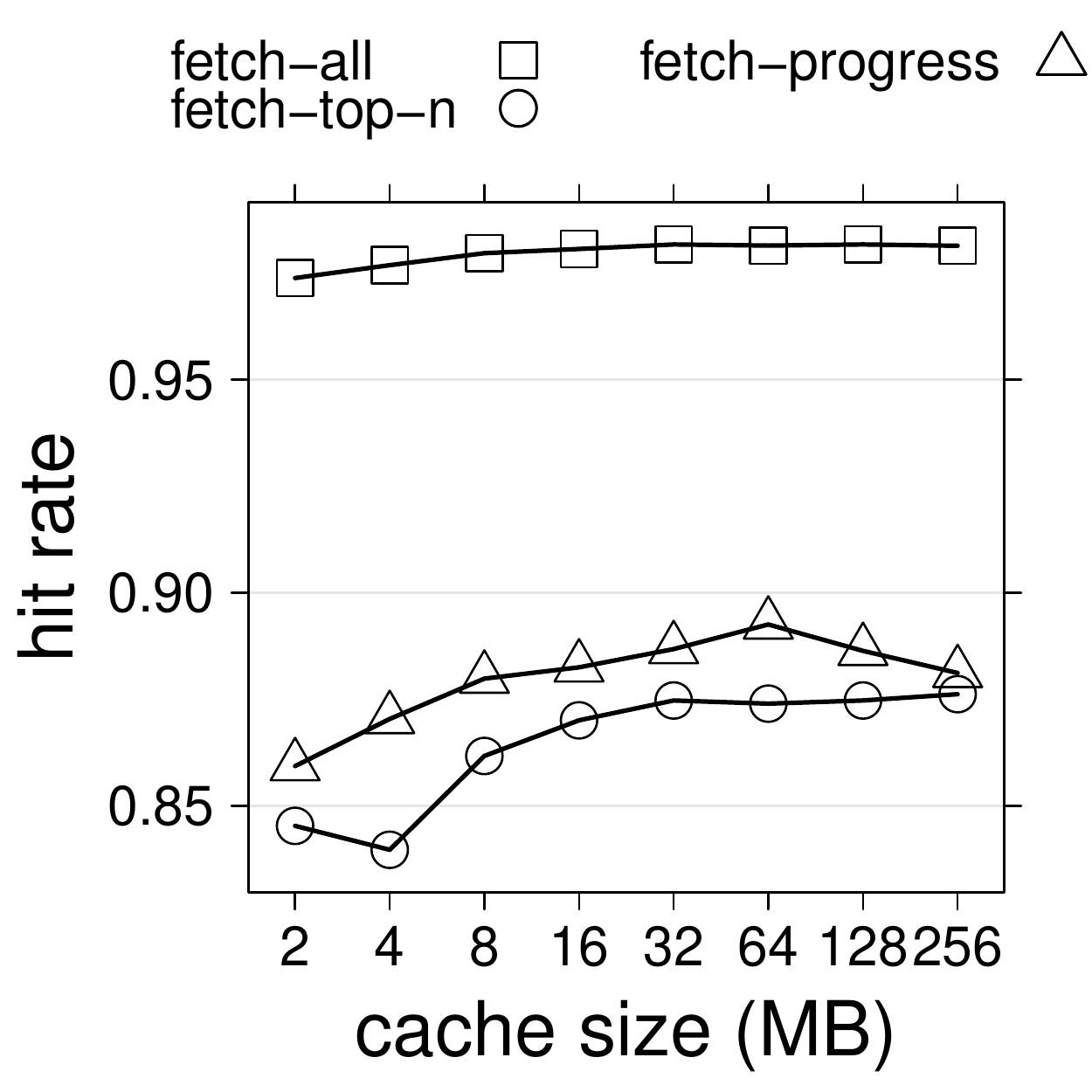}
  	  \caption{Hit rate}
  	  \label{fig:tpcc-hit-rate-cache-size}
  	\end{subfigure} 
  	\begin{subfigure}[t]{0.24\textwidth}
	  \includegraphics[width=\columnwidth]{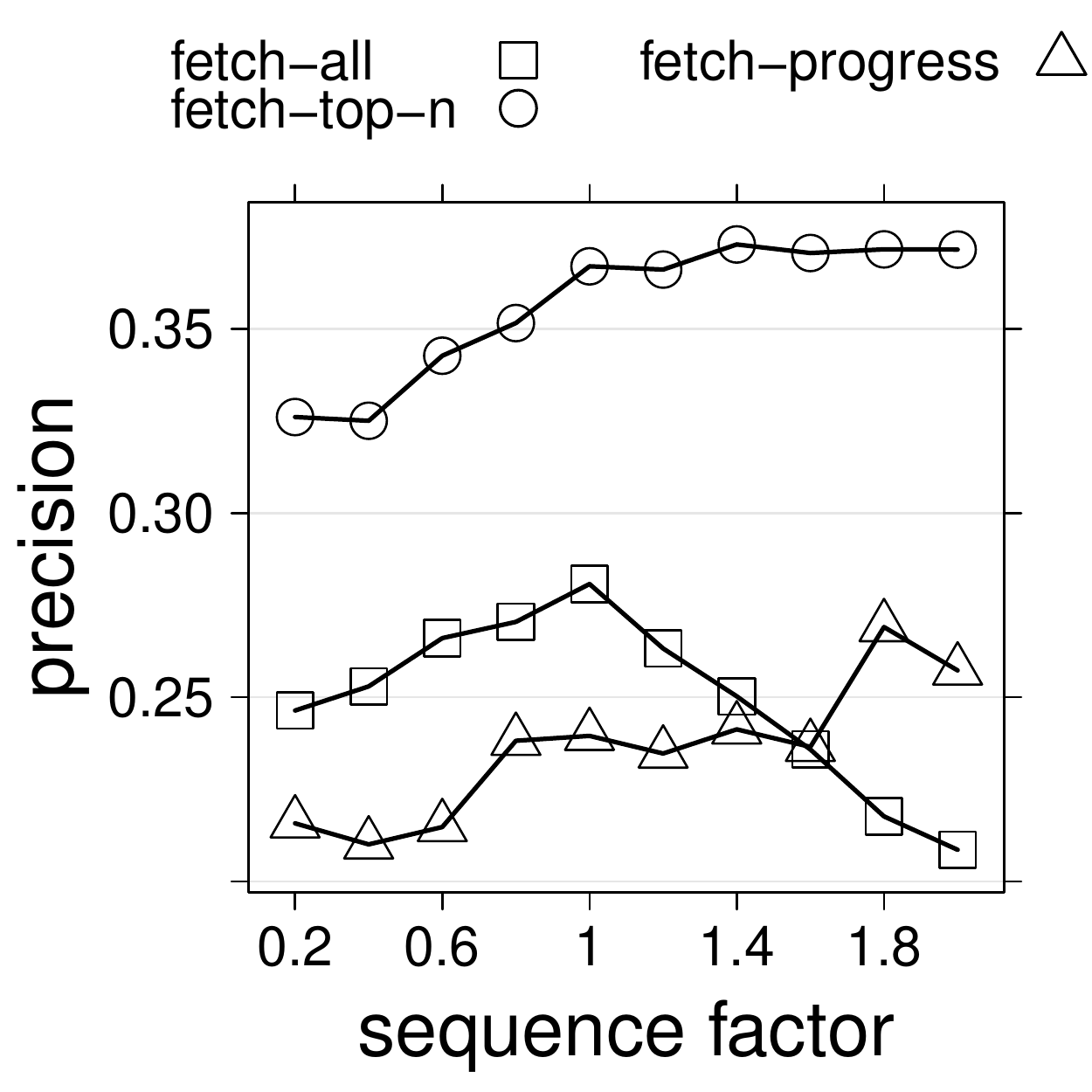}
	  \caption{Precision}
	  \label{fig:tpcc-precision}
	\end{subfigure}
	\begin{subfigure}[t]{0.24\textwidth}
  	  \includegraphics[width=\columnwidth]{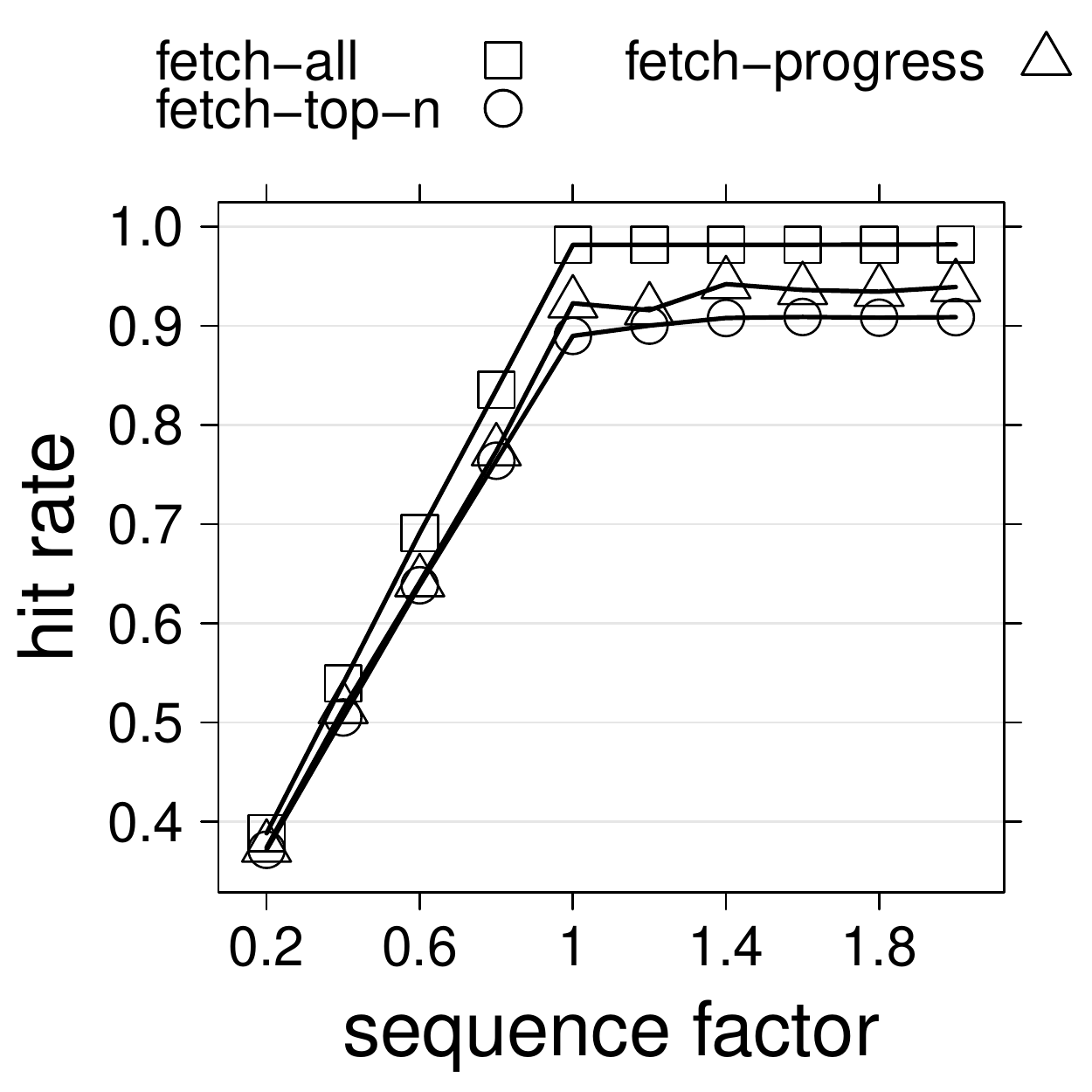}
  	  \caption{Hit rate}
  	  \label{fig:tpcc-hit-rate}
  	\end{subfigure}

	\caption{TPC-C prefetch precision and cache hit rate}
	\label{fig:tpcc-precision-hit-rate}
\end{figure*}

For TPC-C, figures~\ref{fig:tpcc-precision-cache-size} and \ref{fig:tpcc-hit-rate-cache-size} show the impact of the cache size on precision and hit rate for a sequence factor of $1$, that specifies the amount of samples collected from the back log trace (in this case, the amount was of the same size as that of the workload duration). We can see that the precision and hit rate increase with the cache size, albeit not significantly after a size of $32MB$. The obtained hit rate was significantly high, between 80-90\%, across different cache sizes. However, all heuristics, except \emph{fetch-all}, yielded a precision between 20-35\%, which is to expect since TPC-C has a significant non uniform distribution of data accesses.

For the remaining of the experiments, we decided to set the cache size to 32MB since it is a sensible measure (as explained for SEQB) and already provides the most part of the gains in hit rate.

Figures~\ref{fig:tpcc-precision} and \ref{fig:tpcc-hit-rate} illustrate the prefetch precision and hit rate for different heuristics while varying the size of the collected trace (sequence factor) in the first stage. These figures portray the trade-off that we obtain between samples that were traced and accuracy. Like in SEQB, we can observe that  \emph{fetch-top-n} yielded the best precision (between 30-40\%), for the same reasons. The heuristic \emph{fetch-all} had higher precision than \emph{fetch-progress} for most of the sequence factors, meaning that the former performed prefetches that were useful for concurrent accesses. In a workload like TPC-C, where there is not a strong bias on specific sequences, a precision of 30-40\% is significant. Finally, the hit rate increased abruptly with the sequence factor, reaching more than 85\%, and almost 100\% with \emph{fetch-all}, for a sequence factor of at least $1$. These high values were possible to achieve in part due to the \name prefetching mechanisms. Even smaller sequence factors, with hit rates of around 50-60\%, would have significant impact on application latency, performance and web site experience (recall that it results of using just a 32 MB cache).

\begin{figure}
	\centering
	\includegraphics[width=0.9\columnwidth]{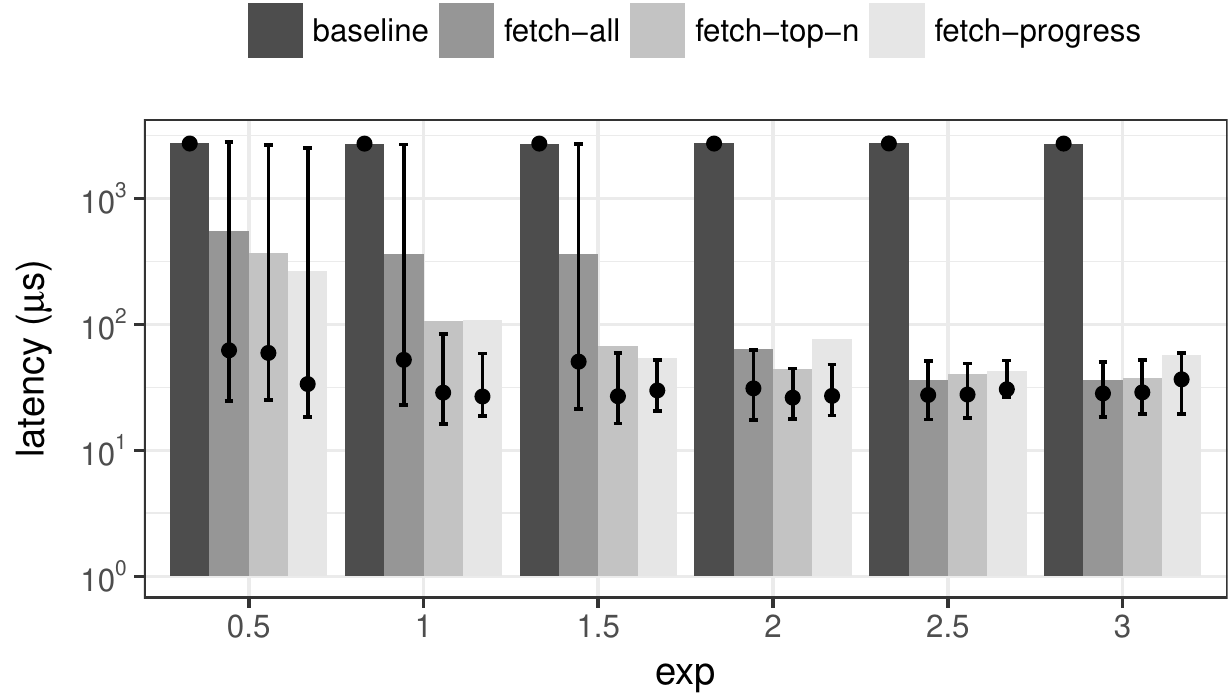}
	\caption{SEQB operation latency. The large bars represent the mean, and the small bars indicate the median (dot), and the 5th and 95th percentiles.}
	\label{fig:seqb-latency}	
\end{figure}


\spara{Latency.} As a consequence of the pronounced hit rate that \name can achieve, we are able to reduce the latency of requests drastically in relation to baseline, as shown in Figure~\ref{fig:seqb-latency} for SEQB. As mean, \name yielded an improvement of 1 up to 2 orders of magnitude with respect to baseline. The \emph{fetch-all} heuristic behavior was more irregular (not decreasing smoothly when exp increases), since a part of its prefetched items were not useful. As median, and without major differences among heuristics, \name had a latency improvement of 75-100x over baseline. The 5th percentile exhibited minimal variation with respect to the median, which indicates that 50\% of the requests were served from \name cache. However, the 95th percentile showed high variation, in relation to the median, when the zipf exponent was $0.5$ for \emph{fetch-top-n} and \emph{fetch-progress}, and lower than $2.0$ for \emph{fetch-all}. This variation indicates that, for less than 45\% of the times, we are not able to anticipate data accesses when there is more entropy in the requests (i.e., prefetching becomes overall less useful). Nonetheless, the latency we obtain with all heuristics is never worse than baseline.

\begin{figure}
	\centering
	\includegraphics[width=0.9\columnwidth]{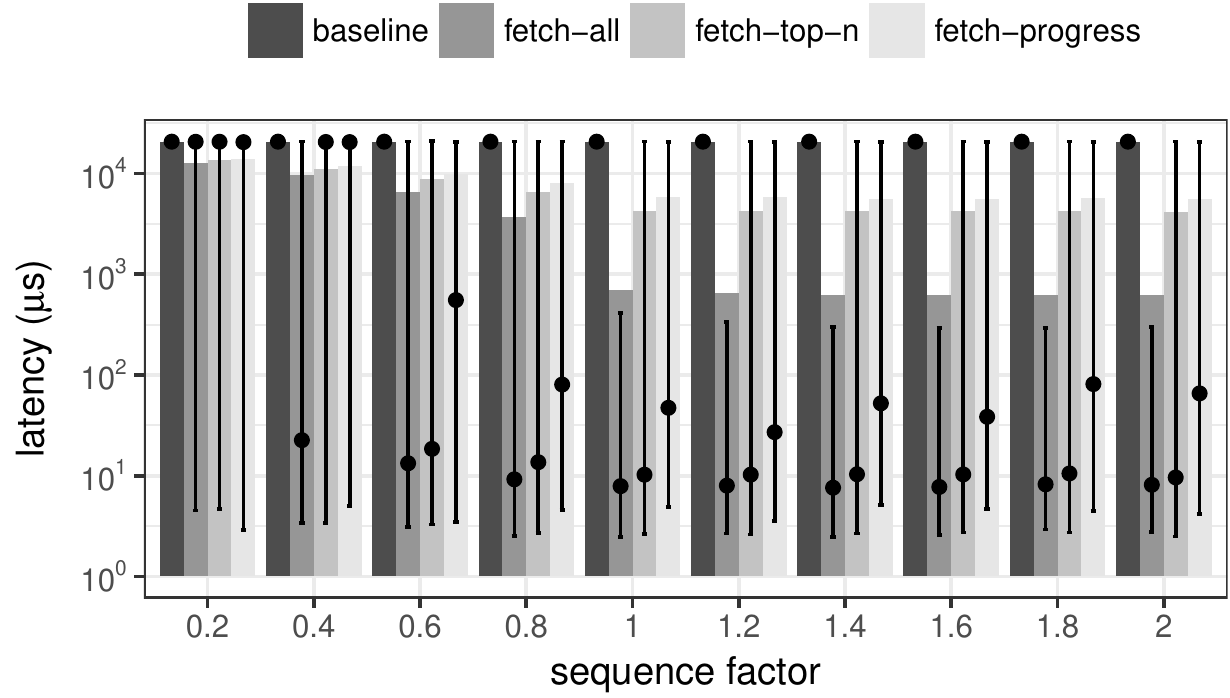}
	\caption{TPC-C operation latency}
	\label{fig:tpcc-latency}
\end{figure}


Figure~\ref{fig:tpcc-latency} shows, for TPC-C, that \name is substantially superior against baseline in terms of latency. On average (mean), \name is 1.5x up to 33x faster than baseline. The heuristic \emph{fetch-all} yielded the lowest latency on average, which comes with no surprise since it: $(i)$ yielded the highest cache hit rate; and $(ii)$ prefetches more items to cache that can also be accessed by concurrent requests. The median values were mostly lower (roughly in one order of magnitude) than the corresponding mean values, indicating that in 50\% of the requests we get a latency that is better than the average request latency. In particular, after a sequence factor of $0.6$, \emph{fetch-all} and \emph{fetch-top-n} achieved peak performance by having a median latency roughly the same as one showed for the 5th percentile. For peak performance (5th percentile), \name is roughly 4000-8000x faster than baseline, which corresponds to serving requests directly from \name cache. Finally, in the worst case (95th percentile), \name was never inferior to baseline, and even exhibited a 70x improvement with the heuristic \emph{fetch-all}.

\begin{figure}
	\centering
	\includegraphics[width=0.9\columnwidth]{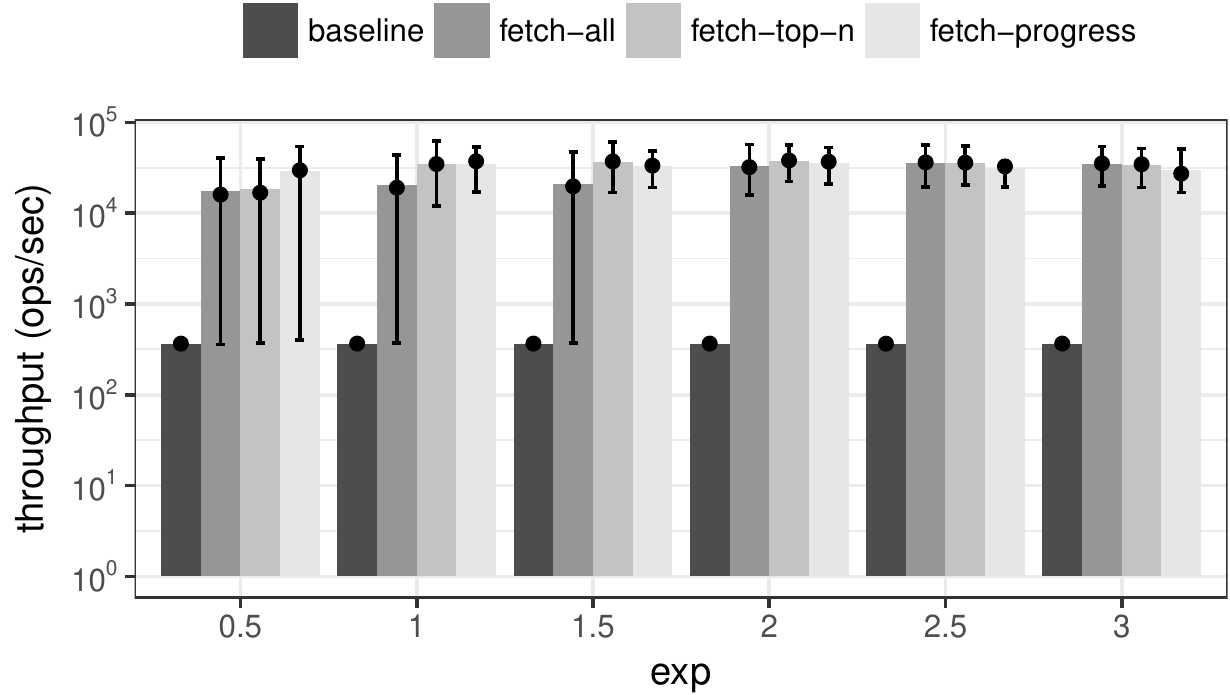}
	\caption{SEQB operation throughput}
	\label{fig:seqb-throughput}	
\end{figure}

\spara{Throughput.} The throughput \name achieved with SEQB is shown in Figure~\ref{fig:seqb-throughput}. On average (mean), \name exhibited a throughput of roughly two orders of magnitude higher than that of baseline for most of the exponents. The median behaved very similar to the mean, indicating that half of the accesses were served from the cache. The 5th percentile shows high variation for \name, especially when the zipf exponent is lower than $2.0$. The explanation for this behavior is the same as before, for the 95th percentile of the latency: more useless prefetching for a small fraction of the times when there are less requests comprising frequent sequences. However, the 95th percentile shows a more stable and high throughput across different exponents. In the most favorable case, \name can process roughly two orders of magnitude more operations that baseline.

\begin{figure}
	\centering
	\includegraphics[width=0.9\columnwidth]{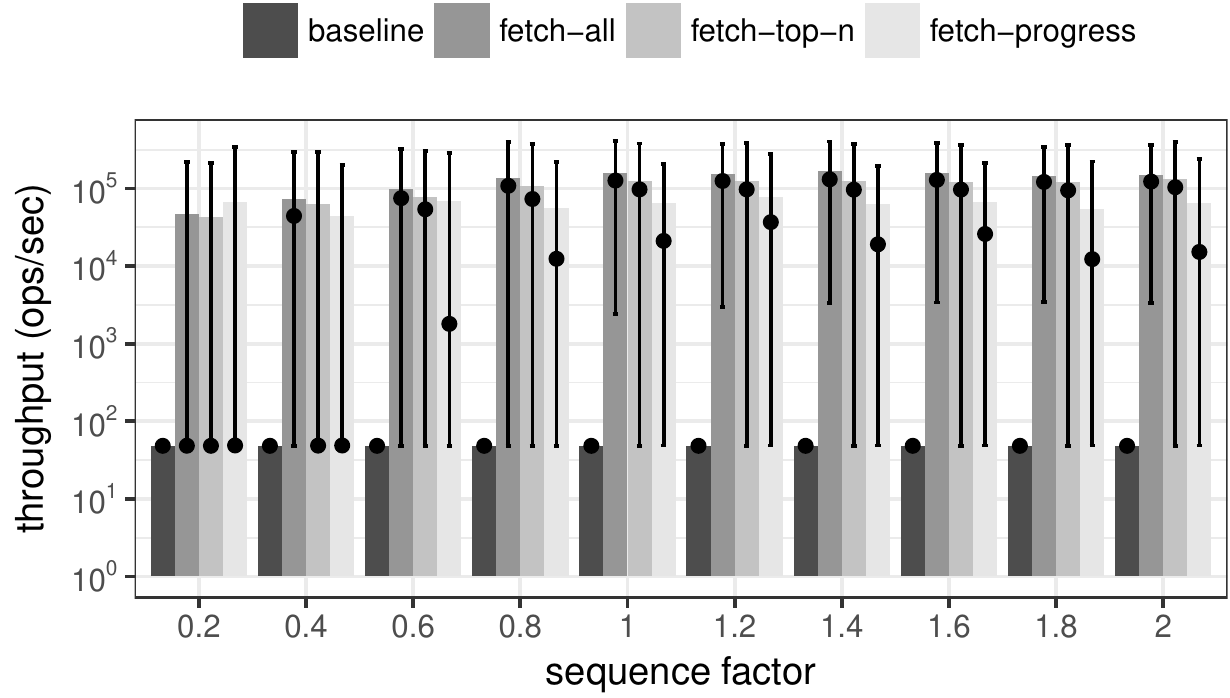}
	\caption{TPC-C operation throughput}
	\label{fig:tpcc-throughput}
\end{figure}

Figure~\ref{fig:tpcc-throughput} shows the throughput obtained for TPC-C. We can observe that, after a sequence factor of $0.4$, the mean and median values are very similar to one another, apart from the \emph{fetch-progressively} heuristic. This means that in 50\% of the times, with a reasonable sequence factor ($>$50\% of the workload size), we get a throughput that is roughly 2-3 orders of magnitude higher than that of baseline. In the worst case (5th percentile), the throughput is not lower to that of baseline and even more 30x higher with the \emph{fetch-all} heuristic for a sequence factor of at least 1. Finally for peak performance (95th percentile), \name can process 3-4 orders more operations per second than baseline.

\begin{figure}
	\includegraphics[width=0.7\columnwidth]{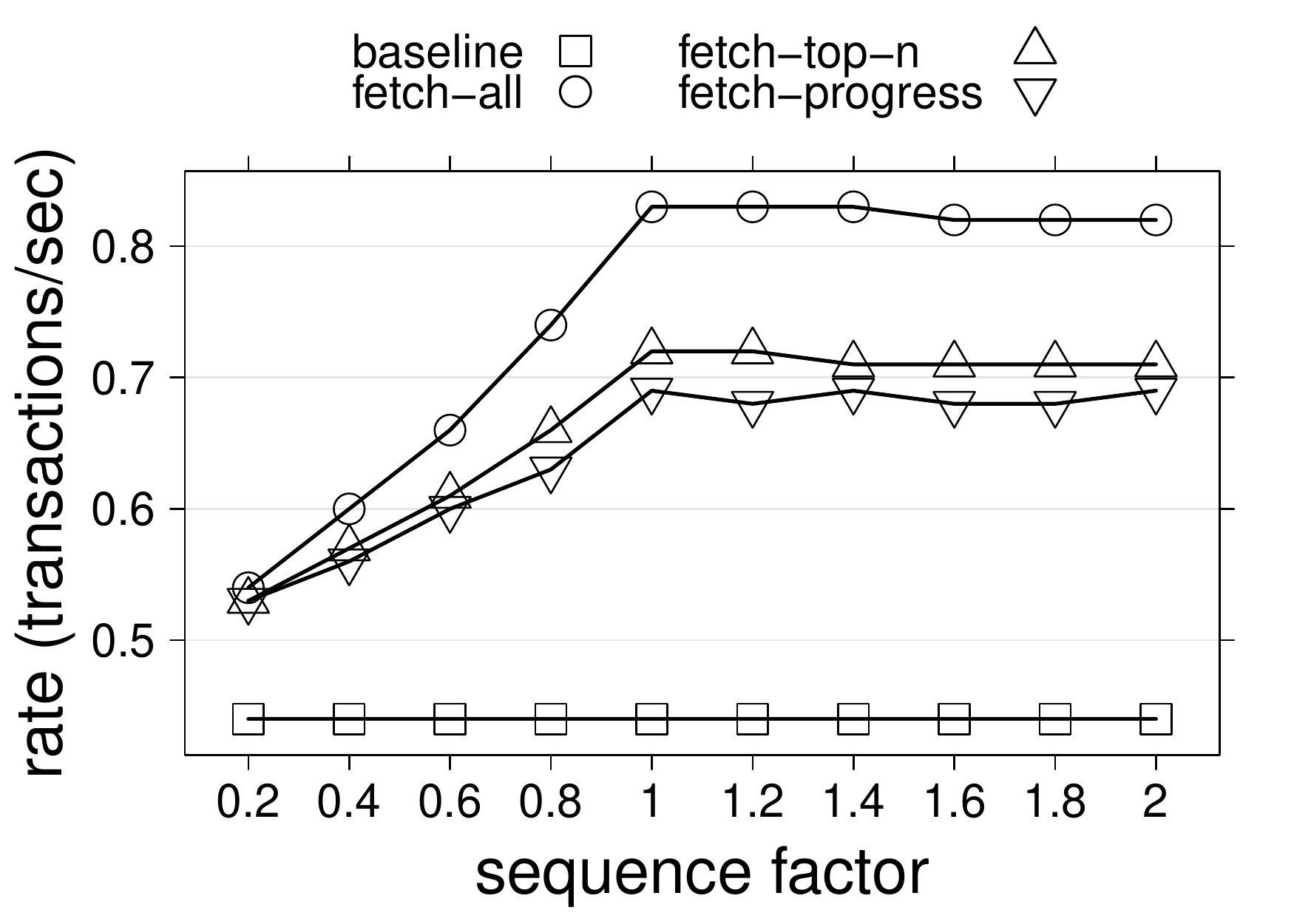}
	\caption{TPC-C rate}
	\label{fig:tpcc-rate}
\end{figure}

In Figure~\ref{fig:tpcc-rate}, we can see the rate of transactions per second (default TPC-C metric). Note that a transaction comprises a large set of read and write operations. \name was able to process 1.2x more transactions than baseline by using a sequence factor of only $0.2$. After collecting a trace equal to the size of the workload ($sequence factor = 1$), \name was almost 2 times better than baseline in terms of transactions processed per second, which is a substantial improvement.

\begin{figure}
	\centering
	\includegraphics[width=0.7\columnwidth]{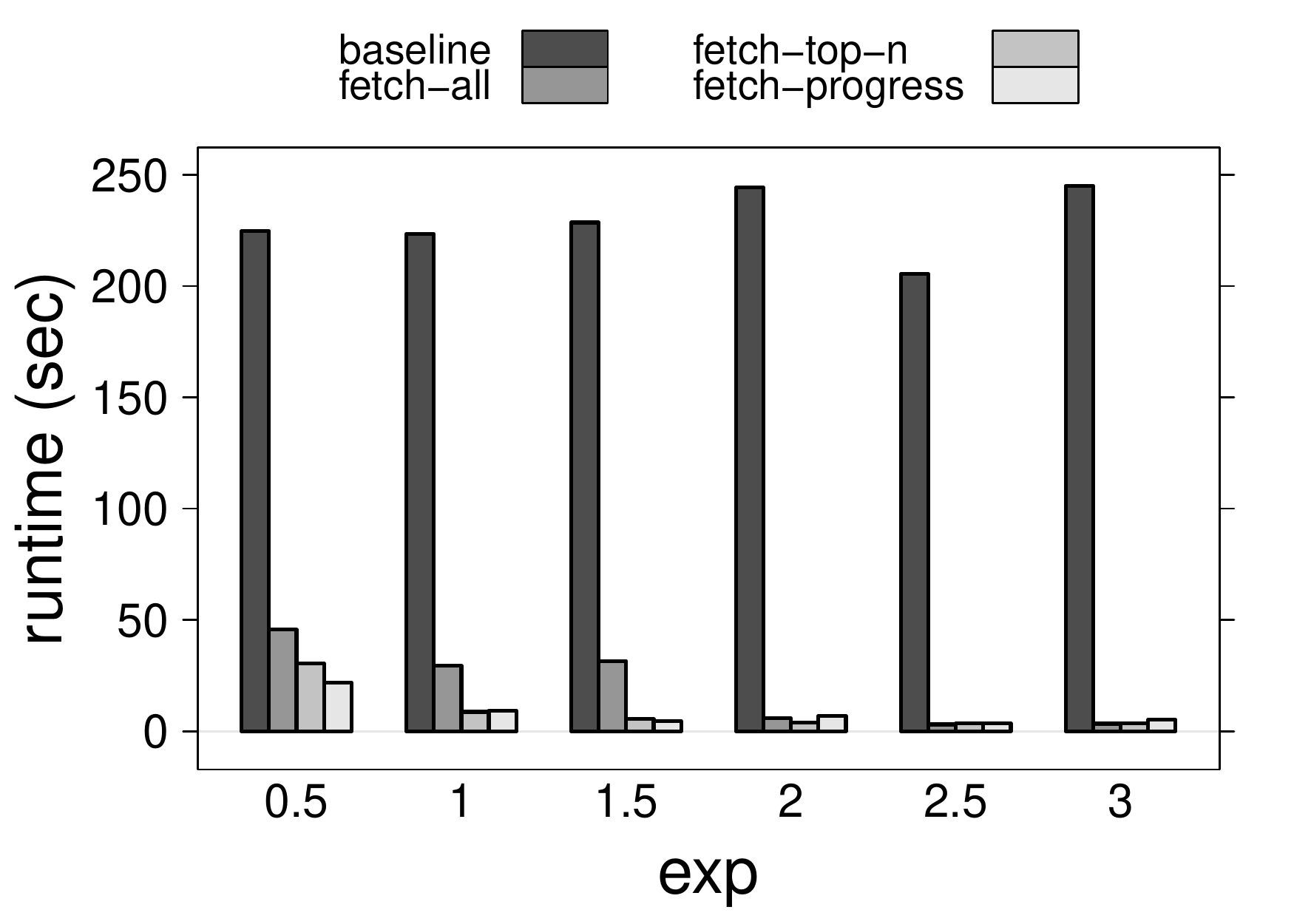}
	\caption{SEQB Runtime}
	\label{fig:seqb-runtime}
\end{figure}

\spara{Runtime.} Figure~\ref{fig:seqb-runtime} depicts the time our workload took to execute for different heuristics and zipf exponents. We can see the execution time for \name heuristics decreases as the exponent increases. This trend is expected since the more operations comprising frequent sequences are executed, the more operations we can anticipate by serving data from the cache. In the less favorable case, \name was about 5x faster than baseline with the \emph{fetch-all} heuristic and exponent of $0.5$; and, in the best case, \name was 66x faster than baseline with also the \emph{fetch-all} heuristic and a exponent of $2.5$.

\begin{figure}
	\includegraphics[width=0.7\columnwidth]{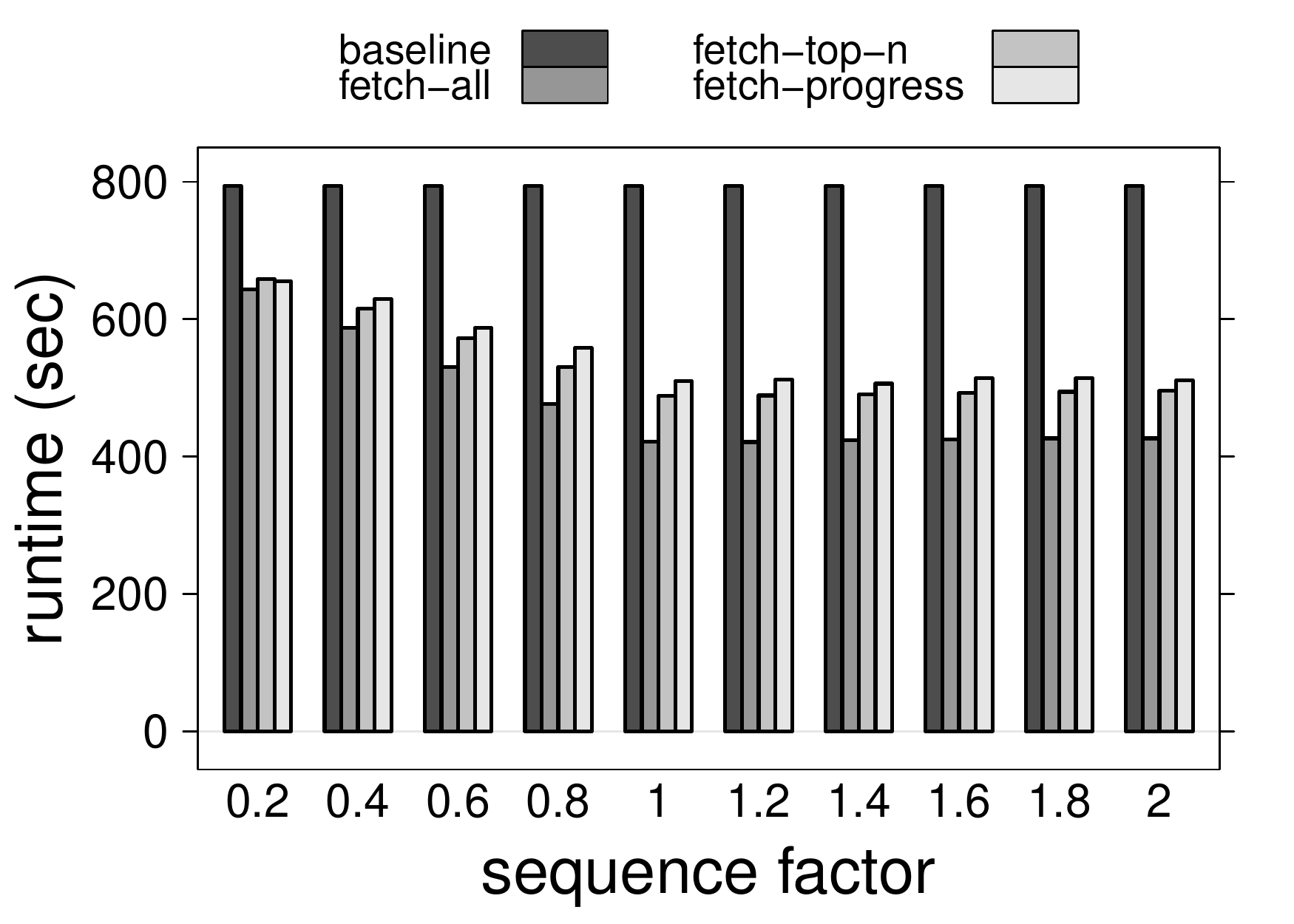}
	\caption{TPC-C Runtime}
	\label{fig:tpcc-runtime}	
\end{figure}

For TPC-C, we can see through Figure~\ref{fig:tpcc-runtime} the time our workload took to execute the 350 pre-defined transactions while varying the amount of observed sequences (sequence factor). Our gains with \name were proportional to the TPC-C rate described before. In the less favorable case, \name completed the execution in about 80\% of the baseline time; and, in the most favorable case, which happened for heuristic \emph{fetch-all} and sequence factor of 1, \name was almost twice faster than baseline. The gains were not higher due to the nature of this benchmark, that performs a mix of reads and writes (i.e., gets, puts and and scans); whereas SEQB is a read-intensive workload, thus yielding higher improvements.

\spara{Reactivity with dynamic workloads.} In a different scenario, that performs online data mining, we assess how \name reacts when the set of frequent sequences changes over time. To this end, we modified the SEQB benchmark to simulate five entirely distinct sets of frequent patterns, referred to as pattern A, B, C, D, and E. Apart from the other configurations, that remain constant, we used the default zipfian exponent (1.0), the fetch-all heuristic, and a cache size that is 33\% the size of that of previous experiments (in order to illustrate our increased benefits with smaller cache sizes). The data mining process is triggered every interval of 20\% of operations for a pattern. The overhead of this process is negligible (less than 1 second of execution time) since we are using a substantially smaller collection of observations than the one described in Section~\ref{sect:mining}.

\begin{figure}
	\centering
	\includegraphics[width=0.9\columnwidth]{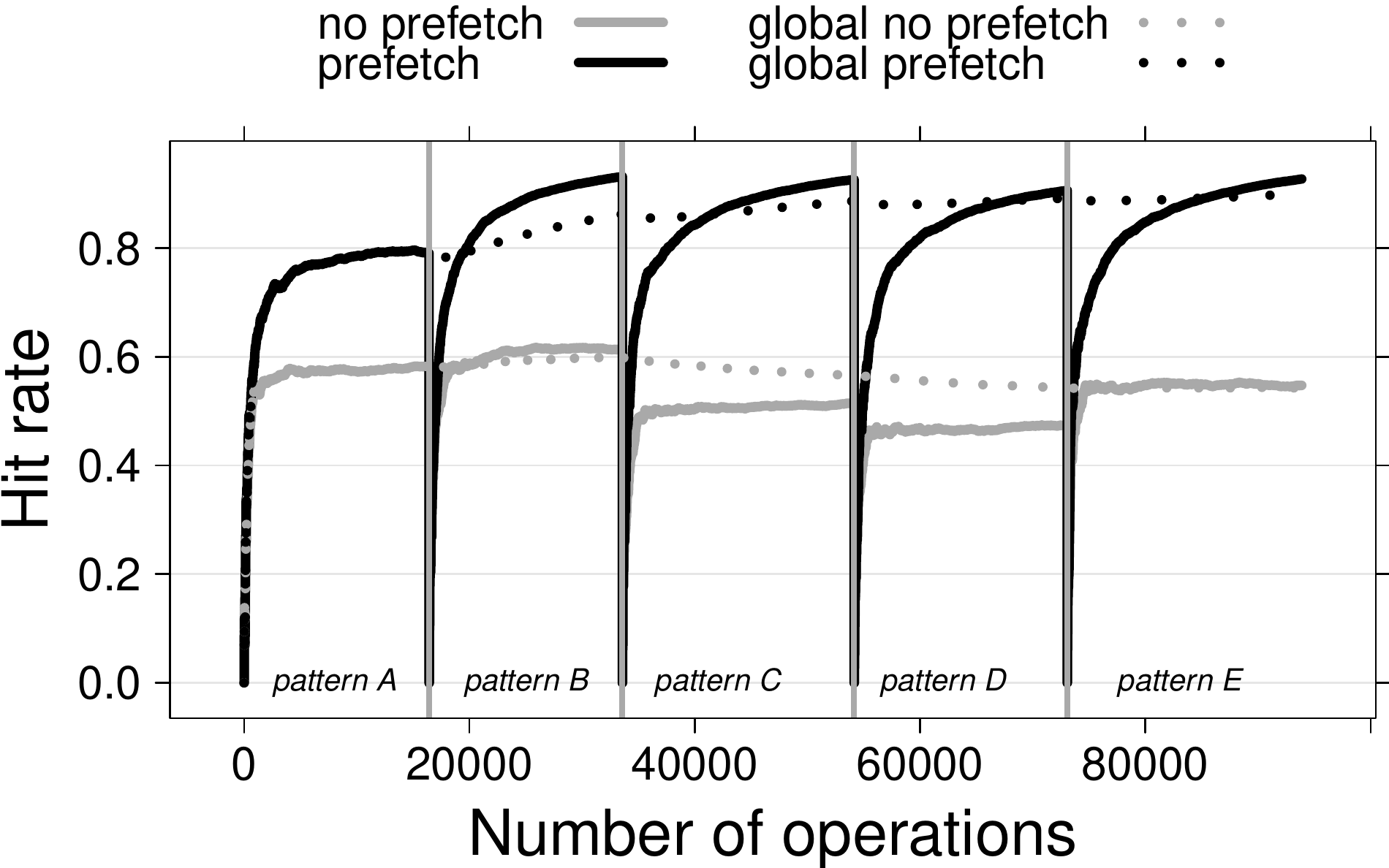}
	\caption{Hit rate when sequence patterns change across time (number of operations). }
	\label{fig:hit-rate-online}
\end{figure}

Figure~\ref{fig:hit-rate-online} shows the hit rate obtained when the sequence patterns change across time (or number of accumulated operations) with prefetching enabled and disabled (i.e., only standard caching). The solid lines show the hit rate locally for each pattern, and the dotted lines represent the global accumulated hit rate since the beginning of time (note the overlapping with the first pattern). Overall, we systematically achieve better hit rate (with long run gains reaching over 30 perc. pts.) with prefetching than solely relying on standard caching. Globally, prefetching recovers faster than standard caching from a drop on hit rate when patterns change; in fact, the global hit rate of just caching keeps decreasing overtime. Further, prefetch continuously improves hit rate as new patterns are observed, whereas standard caching remains at a constant rate and unable to improve much over 60\%.

\spara{Summary.} Overall, SEQB exhibited higher bias and lesser variance than TPC-C in the distribution of data accesses. As a result, SEQB fetched lesser data containers from the back data store, which, by turn, led to higher improvements in terms of runtime (with respect to TPC-C). For the same reason, the heuristic \emph{fetch-all} behaved better in TPC-C than in SEQB; i.e., unlike SEQB, most data prefetched by this heuristic was useful for concurrent and more differentiated accesses in TPC-C.

 
In summary, \name exhibits major performance gains that improve with the recurrence of requests for frequent sequences. These gains position \name as an effective solution to optimize latency in DKV stores, especially when frequent access patterns predominate.

\subsection{Overhead}
\label{sect:overhead}

\begin{figure}
	\centering
	\includegraphics[width=0.7\columnwidth]{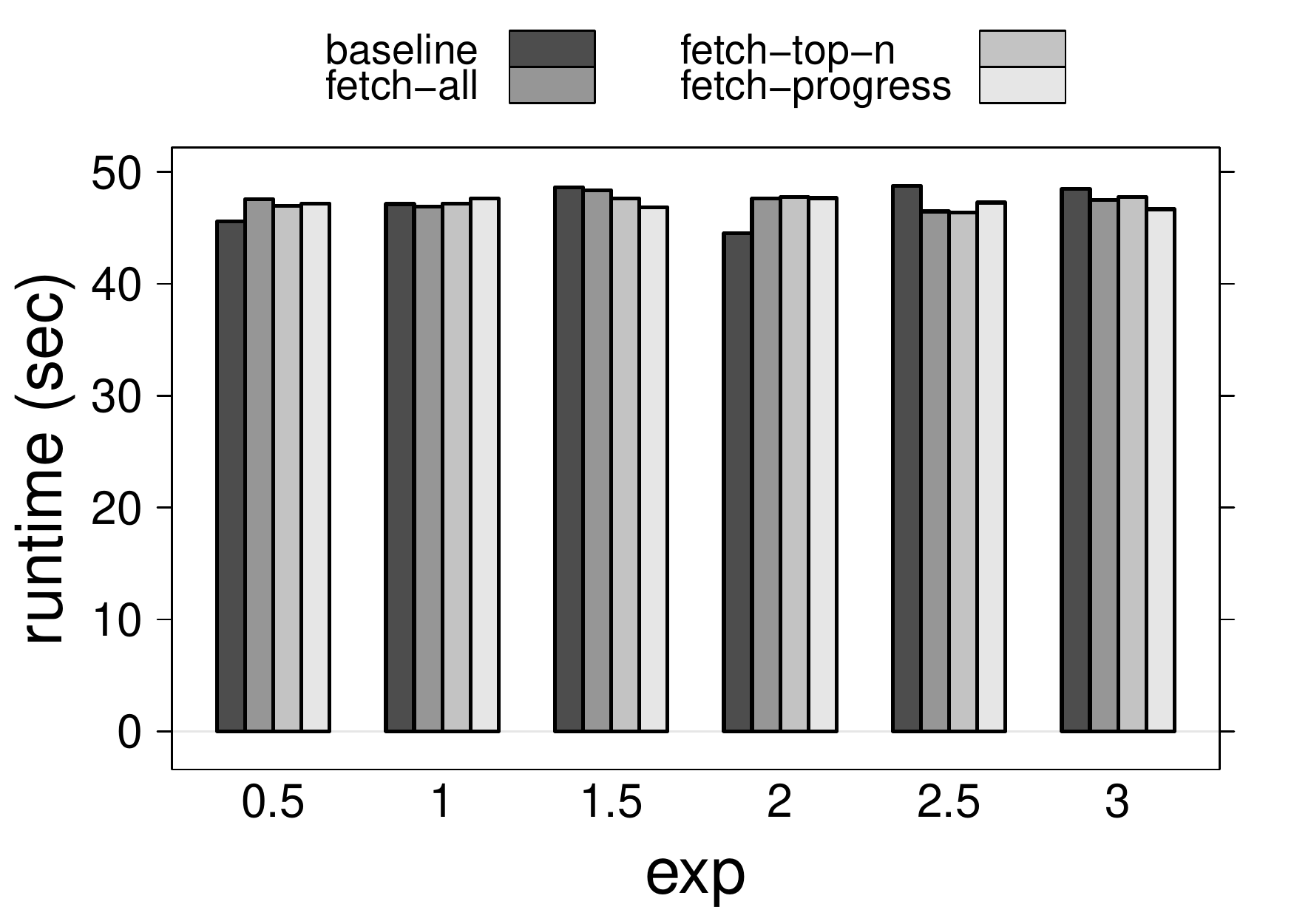}
	\caption{Runtime when \name cache size is set to 0}
	\label{fig:runtime-overhead}
\end{figure}

To measure the overhead of our system, we compared baseline with \name when the cache size is set to 0; i.e., the normal work flow and prefetching mechanisms of \name still take place but without any caching benefits. With a cache of size 0, Figure~\ref{fig:runtime-overhead} shows the runtime for baseline and different heuristics while varying the zipf exponent. We can see that the overall overhead of \name, with any heuristic, is minimal and negligible. In the worst case, \name took 7\% more time to execute than the baseline; and, in the best case, \name took 5\% less time to execute than baseline. This means that most of such variation is noise, and thus the overhead of \name is negligible.

\section{Conclusion}
\label{sect:conclusion}
We presented \name, an in-memory cache at the application level, for DKV stores, that is capable of prefetching data items based on frequently observed patterns. With data mining techniques, \name builds a stochastic graph of frequently accessed sequences of items, and makes prefetching decisions through it.

\name is one of the few systems, if not the first and only, making use of data mining to improve database caching (which can be drastically different from web caching), with special emphasis on caching in NoSQL DKV stores. \name is thus a compelling effort over the state of the art.

Experimental evaluation with realistic benchmarks indicates that \name can improve the latency of HBase by more than an order of magnitude.



\begin{scriptsize}
\bibliographystyle{abbrv}
\bibliography{refs}
\end{scriptsize}

\end{document}